\documentclass[twocolumn,showpacs,preprintnumbers,amsmath,amssymb,prb,aps]{revtex4-2} 
\usepackage{epsfig}
\usepackage{epstopdf}
\usepackage{multirow}
\usepackage{graphicx}
\usepackage{dcolumn}
\usepackage{bm}
\usepackage{textcomp}
\usepackage{array}
\usepackage{csquotes}
\usepackage{subcaption}

\usepackage{booktabs}
\usepackage{gensymb}
\usepackage{amsmath}
\usepackage{bbm}
\usepackage{physics}
\usepackage{hyperref}
\hypersetup{
    colorlinks=true,
    linkcolor=blue,
    filecolor=magenta,      
    urlcolor=blue,
    pdftitle={},
    }
    
\usepackage{titlesec}

\titleformat{\subsubsection}
[block]
{\bfseries\upshape\centering\small} 
{\thesubsubsection.}
{0.5em}
{}

\begin{document}
\title{Unconventional excitations and orbital-driven low-energy dispersions in chiral topological semimetals PdAsS, PdSbSe, and PdBiTe: a \textit{first-principles} study  }
\author{Roopam Pandey}
\altaffiliation{ \url{pandey.roopam09@gmail.com}}

\author{Sudhir K. Pandey}
\altaffiliation{ \url{sudhir@iitmandi.ac.in}}
\affiliation{School of Mechanical and Materials Engineering, Indian Institute of Technology Mandi, Kamand - 175075, India}

\date{\today}

\begin{abstract}

The theoretical dispersion of higher fold excitations are typically governed by space group symmetry. However, physical factors affecting local structural and electronic environment such as atomic arrangement, orbital overlaps, etc., largely alter the behavior of quasiparticle around higher fold nodes. In this work, we consider three chiral material candidates (space group P$2_13$) which exhibit systematic variations in physical parameters by virtue of their constituent elements. We perform a detailed and systematic study of these materials using density functional theory (DFT) in absence and presence of spin-orbit coupling (SOC). Four different kinds of unconventional excitations were observed in all three materials at $\Gamma$- and R-point in the full BZ. In absence of SOC, we find spin-1 ($\Gamma$) and double Weyl (R) excitations, where a Rarita-Schwinger-Weyl fermion ($\Gamma$) and double spin-1 excitation (R) are found in presence of SOC. All of these higher fold nodes lie in energy range of $\left(-0.5,-0.85\right)$eV. Remarkably, we also find total of eight new type-II Weyl points even in absence SOC on $\Gamma$-R line in these materials. In presence of SOC, 12 new Weyl nodes of type-II nature at general momenta ($k_x,k_y,k_z$)$\frac{2\pi}{a}$ are also observed. The presence of these Weyl nodes have not been reported in any of the earlier works. Further, analyzing the low-energy dispersion of spin-1 excitations in these materials we find that otherwise flat middle band in PdBiTe is almost parabolic due strong hybridization. On the other hand, relatively flat middle bands can be observed in PdAsS and PdSbSe in low-energy scale. In case of double spin-1 excitations, surprisingly, we see linearly dispersing middle bands in PdSbSe whereas middle bands in PdAsS and PdSbSe are parabolic even in low-energy scale. We believe our results provide essential insights for realizing and designing topological materials for emerging quantum applications. Lastly, we present non-trivial surface states and Fermi arcs associated with higher fold excitations.

\end{abstract}

\maketitle

\section{Introduction}
Topological semimetals serve as rich platform to explore physics of low-energy quasiparticles extending concepts of relativistic field theory into solid-state physics. The realization Dirac \cite{10.1098/rspa.1928.0023} (Weyl\cite{weyl1929gravitation}) fermions in three dimensional materials opened new avenues of scientific explorations. The electronic structures of both Dirac \cite{PhysRevLett.108.140405,PhysRevB.85.195320,xiong2015evidence,liu2014discovery,liu2014stable} (Weyl \cite{PhysRevB.83.205101,lv2015observation, PhysRevX.5.031013,xu2015discovery,xu2015discoverys}) semimetals feature distinct points of fourfold (twofold) degeneracy in the Brillouin zone (BZ) with bands dispersing linearly in the momentum space around the band touching points generically known as nodes \cite{RevModPhys.90.015001}. Weyl semimetals, especially, have attracted considerable theoretical and experimental interest due to the unique role of their nodes as monopoles of Berry curvature; where each Weyl node carries a monopole/topological charge, C$=\pm1$. In the vicinity of a Weyl node, fermions behave as if subjected to an effective magnetic monopole field generated by the Berry curvature \cite{RevModPhys.90.015001}. Weyl semimetals are known to exhibit fascinating Berry curvature driven phenomena such as chiral anomaly \cite{PhysRevB.86.115133,PhysRevLett.117.146603,RevModPhys.90.015001}, Weyl fermions quantum transport and Hall effects \cite{PhysRevLett.111.027201,PhysRevLett.113.187202, PhysRevLett.117.146403,RevModPhys.90.015001}. These semimetals also demonstrate an atypical and quantized responses to light \cite{de2017quantized,PhysRevB.98.165113,wu2017giant}. 

While Poincar\'e symmetries imposed stringent constraints on type of particles that might occur in high-energy physics, condensed matter systems provided more flexible grounds for discovery of fermionic quasiparticles which had no high-energy counterparts. Quasiparticles in solids only need to adhere to crystal symmetries of the space group (SG) systems (230 SGs). Recent advancements have identified unconventional quasiparticles and higher band degeneracies different than those found in standard topological semimetals \textit{viz.} three-, four-, six-, and eightfold crossings which have no high-energy analogs \cite{bradlyn2016beyond}. By the virtue of crystal symmetries, band theory in solids allows for higher fold degeneracies to typically appear at time-reversal invariant momenta (TRIM, $\vec{\textbf{\textit{k}}}$) in some of the SGs out of 230 SGs \cite{bradlyn2016beyond,bradley2009mathematical}. The degeneracy of these higher fold nodes at $\vec{\textbf{\textit{k}}}$ is governed by dimension of irreducible representations (irreps) of the little group at $\vec{\textbf{\textit{k}}}$. Further, these higher fold nodes carry topological charges greater that C$=\lvert\pm1\rvert$ i.e., 2 and 4; except for eightfold degeneracy which has C$=0$ \cite{bradlyn2016beyond}.

Having established that crystallographic symmetries can enforce higher fold topological nodes, we now turn our attention to emergent phenomena in topological semimetals due their structural chirality. In solid-state context, crystal structures that lack both inversion and mirror symmetries are known as chiral crystals. Topological semimetals that possess chiral structures are referred as chiral topological semimetals \cite{schroter2019chiral, chang2018topological}. 
Chiral topological semimetals are anticipated to exhibit a range of unique physical phenomena including fermionic excitations carrying substantial topological charge \cite{bradlyn2016beyond,schroter2019chiral} and associated Fermi arc surface states extending throughout the BZ \cite{PhysRevLett.119.206401}.
Further, owing to large berry curvature due to large topological charges atypical magnetotransport \cite{zhong2016gyrotropic,PhysRevB.108.035428} and lattice dynamics \cite{rinkel2017signatures} are also observed. Structural chirality along with non-trivial topological features confers these systems with a precise quantized response to circularly polarized light \cite{schroter2019chiral,de2017quantized, PhysRevB.98.155145,ni2021giant}. These exotic phenomena can potentially find applications in spintronics \cite{shi2019all,hirohata2020review}, next generation optoelectronic devices \cite{ma2017direct,he2024selective}, chemical catalysis \cite{rajamathi2017weyl}, etc. The aforementioned intriguing and measurable physical phenomena have thus led to grown interest in finding novel chiral topological semimetallic candidate materials with higher fold excitations.

Until now, three-, four-, and sixfold degenarcies have been realized in topological chiral crystals. Low-energy behavior of a threefold denegracy can be described as spin-1 excitations, carrying monopole charge, C=$\pm2$ \cite{bradlyn2016beyond}. Sixfold degeneracy which is essentially two spin-1 excitations stabilized at same momenta in these systems carrying a topological charge of $\pm4$ \cite{bradlyn2016beyond, PhysRevB.102.155147, PhysRevB.99.241104}. Further, two distinct type of fourfold degeneracies are observed namely, Double Weyl (charge-2) \cite{PhysRevLett.119.206402,PhysRevB.99.241104} and Rarita-Schwinger-Weyl (spin-3/2) fermion \cite{PhysRev.60.61} (RSWF/RSWP) . A Double Weyl point has C=$\pm2$ whereas RSWF has C=$\pm4$ .

Previously, family of rare-earth metal carbides \cite{PhysRevB.104.045111}, transition metal silicides \cite{PhysRevLett.119.206402}, and electrides \cite{PhysRevResearch.3.L012028} have been studied to explore the higher fold excitations. Transition metal silicides such as CoSi \cite{rao2019observation,sanchez2019topological,PhysRevLett.122.076402,PhysRevLett.129.026401,pshenay2018band} and RhSi \cite{PhysRevLett.119.206402,chang2017unconventional,rees2020helicity,maulana2020optical}, and AlPt \cite{schroter2019chiral} have been studied theoretically and have also been probed for the non-linear optical response experimentally \cite{PhysRevLett.119.206401,ni2021giant}. Few ternary and quaternary systems like PdBiSe \cite{PhysRevB.99.241104}, PdSbSe \cite{PhysRevMaterials.9.L031201} and KMgBO$_3$ \cite{PhysRevB.102.155147} have also been studied. Further, establishing their non-trivial toplogical character many of these materials host Fermi arcs spanning the whole BZ. Specifically in CoSi, RhSi, and AlPt these long Fermi arcs have been identified experimentally as well \cite{rao2019observation,sanchez2019topological,chang2017unconventional,schroter2019chiral}. 

Among these materials PdBiSe \cite{PhysRevB.99.241104} and PdSbSe \cite{PhysRevMaterials.9.L031201} are relatively less explored, but may prove to be of great interests as PdBiSe is a known superconductor \cite{joshi2015superconductivity}. Investigating their topological properties will further widen their scope of applications. Previous studies on PdBiSe and PdSbSe verify the presence of higher fold nodes at high symmetry points via angle resolved photoemission spectroscopy (ARPES) experiments. Further, analysis of bulk electronic structure of these materials from DFT is limited to high symmtery points and lines where degeneracies are symmetry enforced and their appearance is certain. However, it is equally important to account for degeneracies which might occur at momenta other than high symmtery points and lines requiring a thorough study of the full BZ. Therefore, we present an extensive and exploratory study of bulk electronic structure using DFT in the full BZ for PdSbSe and its two isostructural compounds in this work.

Although, the higher fold nodes originate from space group symmetry which ensures and determines the degeneracy at a point, the actual behavior of quasiparticles in the vicinity of these nodes is determined by local electronic and structural environment. 
Ideally, low-energy behavior of a quasiparticle in neighborhood of $\vec{\textbf{\textit{k}}}$ can fairly be described by a linearized $\vec{\textbf{\textit{k}}}\cdot\vec{\textbf{\textit{p}}}$ Hamiltonian which complies with symmetries of little group at $\vec{\textbf{\textit{k}}}$. These $\vec{\textbf{\textit{k}}}\cdot\vec{\textbf{\textit{p}}}$ Hamiltonian result in dispersions that are characteristic to the unconventional excitations. For example low-energy dispersion of spin-1 excitation features two linearly dispersing bands with a middle flat band. Consequently, we examine how physical factors such as arrangement of atoms, overlap of orbitals, orbital characters, strength of SOC, etc., might shape the low-energy dispersion of quasiparticles in real materials. Accordingly, we take three materials namely, PdAsS, PdSbSe, and PdBiTe well-suited for objectives our study. All three materials belong to chiral space group P$2_13$. Owing to their constituent elements, these materials exhibit systematically varying atomic radii, spatial expanse of orbitals, and strength of SOC. 

We begin our study with examining their bulk electronic structure using exploratory DFT calculations to identify and locate unconventional, Weyl or Dirac crossings in the full BZ in presence and absence of SOC. Interestingly, we find eight Weyl points even in absence of SOC and 12 new Weyl points at general momenta ($k_x,k_y,k_z$)$\frac{2\pi}{a}$ in presence of SOC. These results have not been reported in any of the available works. Further to gain insight into low-energy physics of the unconventional quasiparticles in these materials, we plot low-energy dispersion around all node points. Orbital characters are also analyzed to correlate deviation in these dispersion and role of hybridization. Notable deviations were observed in ideally flat middle band(s) of spin-1 and double spin-1 excitations. Low-energy dispersion of spin-1 excitation in PdBiTe features a middle parabolic band, whereas in PdAsS and PdSbSe it remains flat within a small region around $\Gamma$. Further, in presence of SOC, we find rather linearly dispersing middle bands which otherwise are flat in PdSbSe, whereas in PdAsS and PdBiTe these bands are parabolic even in low-energy region. We also plot dispersions around obtained Weyl points to establish their type-II nature. Chirality of all bands forming topological nodes have been calculated and presented in Table \ref{tabb} for higher fold nodes. Finally, we plot surface spectra and examine non-trivial surface states and Fermi arcs associated with higher fold excitations. 


\section{Computational details}
The \textit{first-principles} calculations were performed using Density Functional theory (DFT) with augmented -plane wave plus local orbital (APW+lo) basis set as implemented in the WIEN2k package \cite{blaha2020wien2k}. The Perdew-Burke-Ernzerhof for solids (PBEsol) \cite{PhysRevLett.100.136406} exchange-correlational functional was used for the electronic structure calculations. The muffin-tin sphere radii (\textit{$R_{MT}$}) for all elements of systems studied here are given in Table ST1 of \cite{supp}. The ground state calculations were performed on a \textit{k}-mesh of size 20$\times$20$\times$20 (sampled in first Brillouin zone) with a energy convergence of $10^{-7} $ Ry/cell. \textit{PY-Nodes} \cite{PANDEY2023108570} code, based on \textit{Nelder-Mead's function-minimization} method, was used to find the nodes.
In order to study topological signatures i.e, non-trivial surface states and Fermi arcs, the Tight-Binding(TB) scheme was used to obtain maximally-localised wannier functions (MLWFs) in presence of SOC using Wannier90 \cite{Pizzi2020} code. MLWFs were contructed using \textit{p-}(\textit{As, Sb,} and \textit{Bi}) and \textit{d-}( \textit{Pd}) orbitals. Further, this TB model is used to calculate the surface spectra and arcs using the iterative Green function method as implemented in WANNIERTOOLS \cite{WU2018405} code. The topological invariants were calculated using WloopPHI \cite{SAINI2022108147}. The optimized lattice parameters ($\textit{a}$) and atomic positions of PdAsS, PdBiTe, and PdSbSe are given in Table ST1 of \cite{supp}.


\section{Results and Discussion}
\subsection{Electronic Structure}

\begin{figure}[t]
	\centering
	\includegraphics[width=0.498\textwidth]{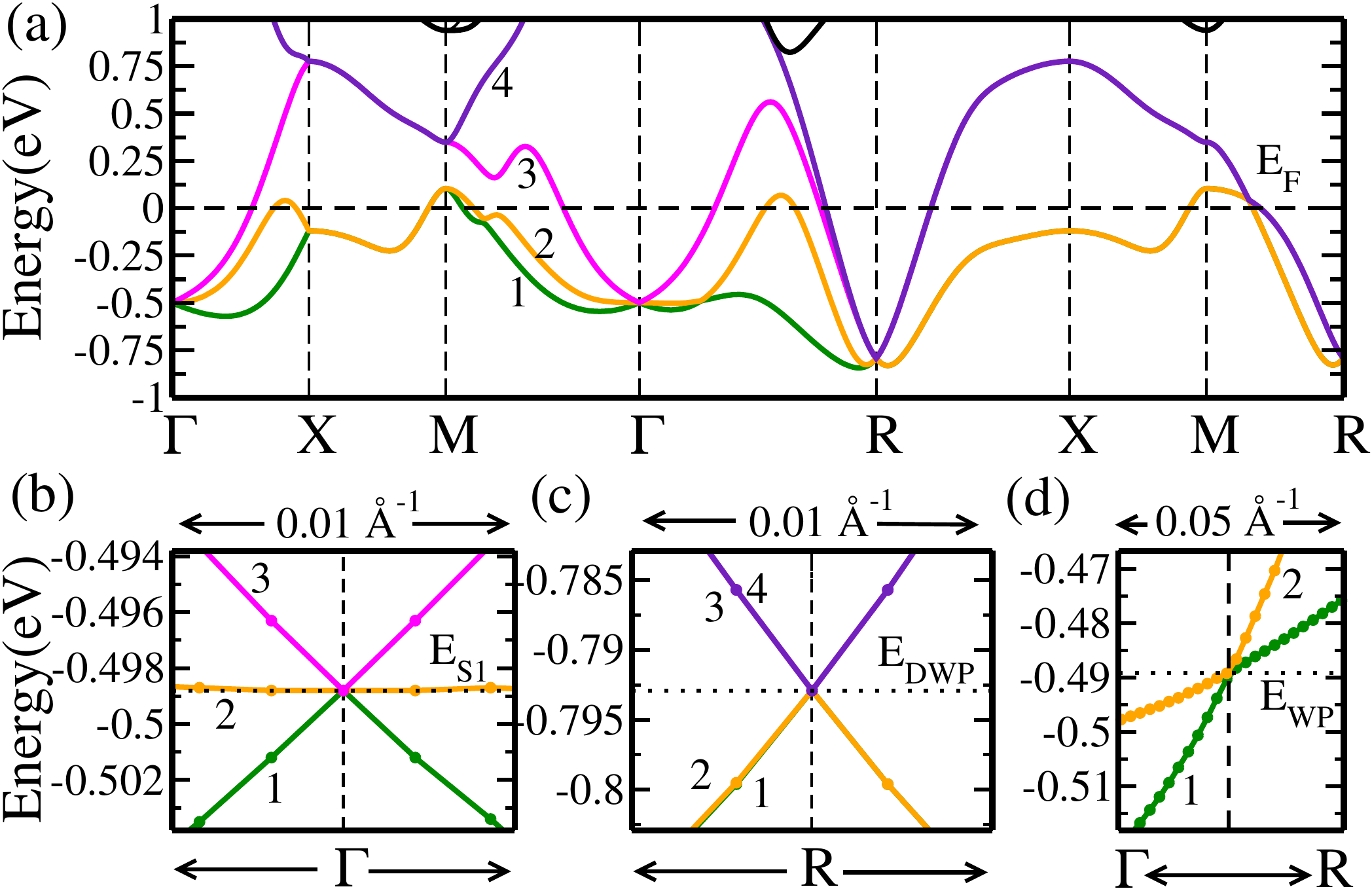}
	\caption{Electronic band structures without spin-orbit coupling (SOC). (a) Full dispersion of PdAsS with line E$_F$ representing Fermi level; Zoomed-in view (b) at $\Gamma$-point showing spin-1 excitation, where E$_{S1}$ is energy of the three fold degenerate state; (c ) a charge two fourfold fermion at R-point and energy E$_{DWP}$ and (d) showing type-II Weyl point along $\Gamma$-R with energy E$_{WP}$ . }
	\label{fig:nsoc}
\end{figure}

    \begin{figure}[t]
	\captionsetup[subfigure]{labelformat=empty}
	\centering
	\begin{subfigure}[b]{0.15\textwidth}            
		\includegraphics[width=\textwidth,height=3cm]{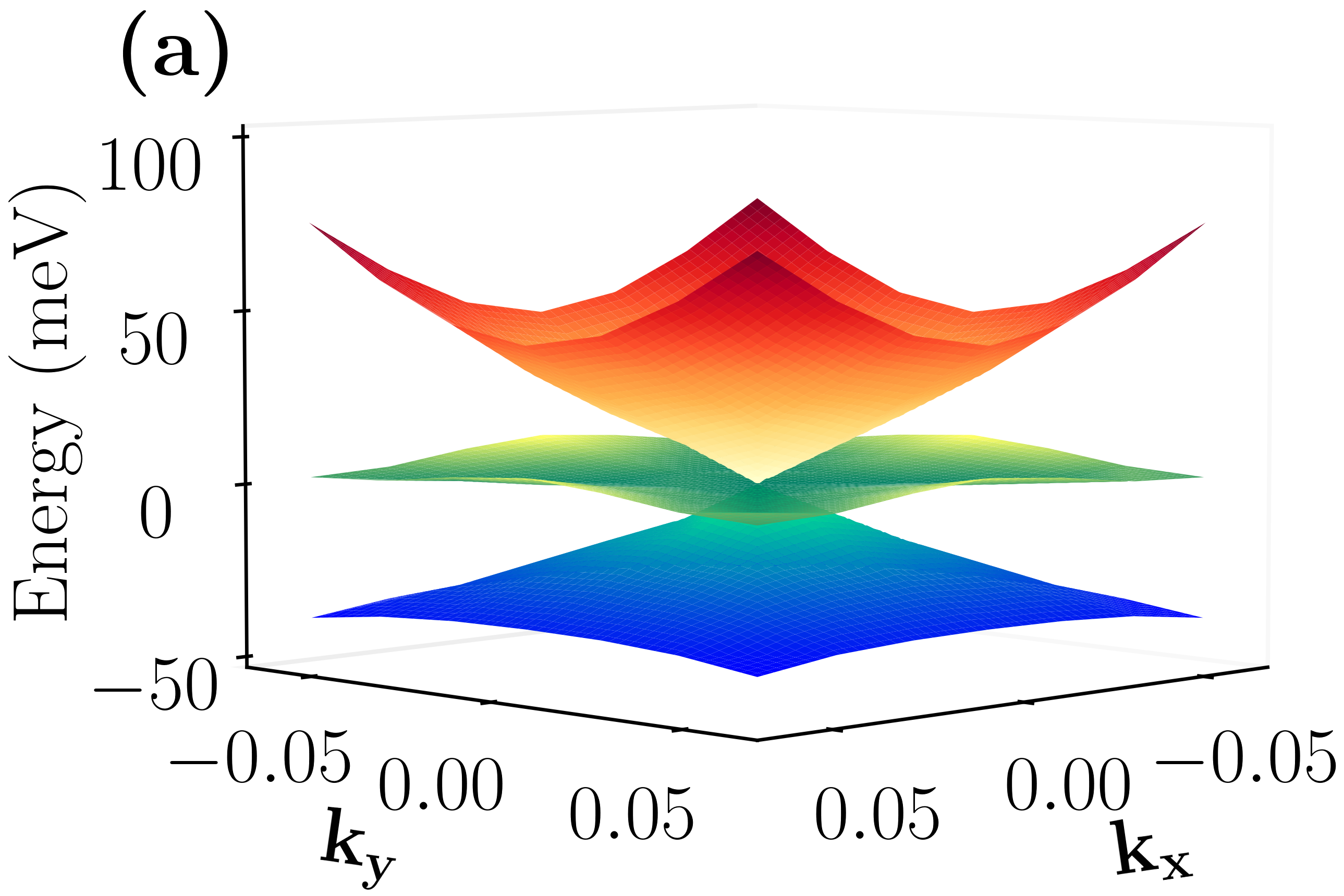}
		\caption{\hspace{-1em}}		
	\end{subfigure}%
	\begin{subfigure}[b]{0.16\textwidth}
		\centering
		\includegraphics[width=\textwidth,height=3cm]{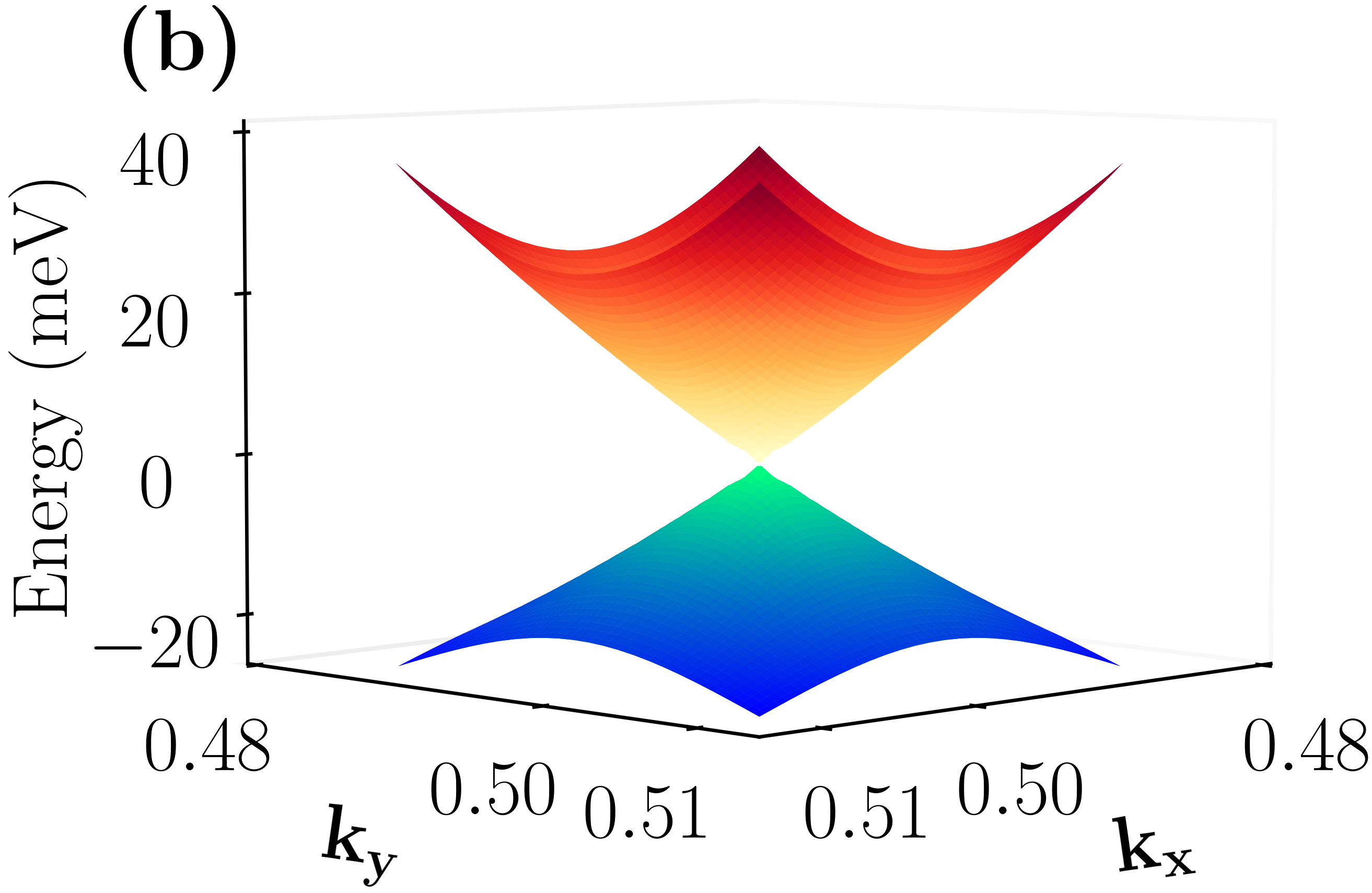}
		\caption{\hspace{-1em}}		
	\end{subfigure}
	\begin{subfigure}[b]{0.16\textwidth}            
		\includegraphics[width=\textwidth,height=3cm]{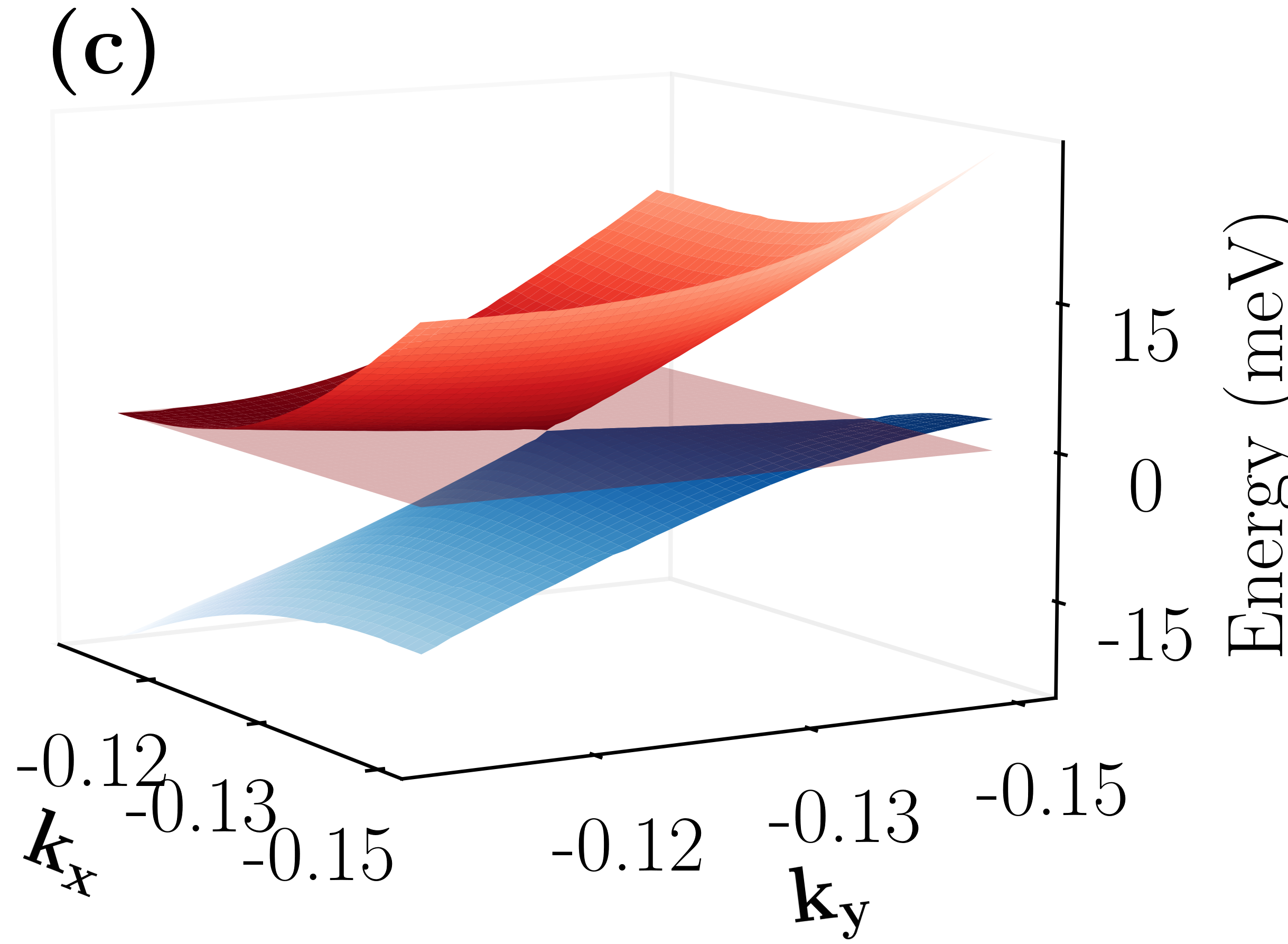}
		\caption{\hspace{-1em}}		
	\end{subfigure}%

	\caption{The figure displays energy dispersion in xy-plane near the multifold nodes at $\Gamma$- and R- point in the BZ of PdAsS. (a) spin-1 excitation ; (b) Charge-2 fourfold fermion multifold; (c) type-II Weyl point in $\Gamma$-R direction. z-axis represnts energy in meV. Here, the energy of nodes is scaled to zero in all the subfigures. }\label{fig:ndisp}
\end{figure}


\begin{figure}[t]
	\centering
	\includegraphics[width=0.48\textwidth]{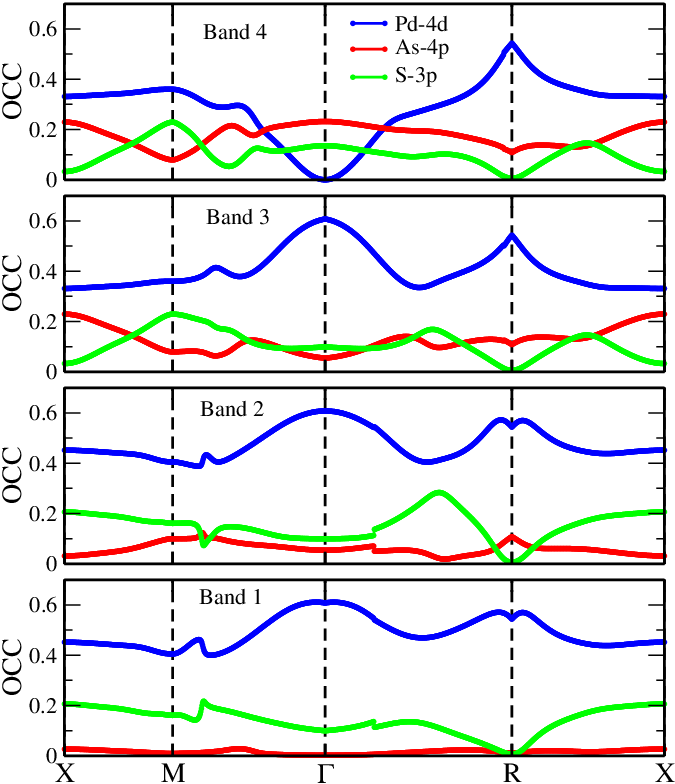}
	\caption{ Orbital character contributions to the bands forming multifold nodes without SOC. Band 1-3 form spin-1 excitation at $\Gamma$ and Band 1-4 double Weyl point at R}
	\label{fig:nocc}
\end{figure}

 In the present study we have performed exploratory study on three material namely PdAsS, PdSbSe, and PdBiTe. All the three materials taken into consideration here belong to chiral space group P2$_1$3. It is known that materials belonging to this class of compounds exhibit multifold excitations at high-symmetry points. In this section we aim to corroborate the established results and also to study any new crossing/excitations that might occur in these system. Along with this, it is also intended to study how these excitations vary across the materials. Electronic structures of all the materials were calculated without SOC in high symmetric direction $\Gamma-X-M-\Gamma-R-X-M-R$, and a full band structure of PdAsS is presented in Fig. \ref{fig:nsoc} (a) with Fermi level (E$_F$) set to zero. The band structures for PdSbSe and PdBiTe can be found in Sec. IIA (Fig. S1 (a) and (e)) of supplementary material \cite{supp}. It is typically seen that materials belonging to this class of compound host a spin-1 excitation and double-Weyl or charge-2 fourfold fermion at $\Gamma$- and R-point, respectively \cite{bradlyn2016beyond,PhysRevB.99.241104}. Our calculations replicate the results consistent with previously reported studies in all the three materials. Fig. \ref{fig:nsoc}(b) and (c) show zoomed in view of spin-1 excitation at $\Gamma$- (bands 1-3, $\sim$-0.5 eV (E$_{S1}$)) and a fourfold degenerate state (bands 1-4, $\sim$-0.8 eV (E$_{DWP}$)) at R-point in PdAsS. Spin-1 excitation in PdSbSe and PdBiTe occur at energies $\sim$-0.85 eV and $\sim$-0.78 eV, respectively. The dispersion around point-R (Fig. \ref{fig:nsoc} (c)) across the materials also show the respective change in slope around R from PdAsS (shallow) to PdBiTe (steep). Meanwhile the fourfold degenerate node in both PdSbSe and PdBiTe are located in energy window -0.8 to -0.9 eV. Interestingly, our calculations also indicate presence of type-II Weyl points along the $\Gamma$–R direction in all three materials, even in the absence of SOC $-$a phenomenon typically associated with SOC effects. Two sets of Weyl points were identified (WP1 $\&$ WP2), each containing four points with representative coordinates ($k_0,k_0,k_0$)$\frac{2\pi}{a}$ and ($-k_0,-k_0,-k_0$)$\frac{2\pi}{a}$. The calculations explicity revealed three (WP1) and four (WP2) points, and rest were obtained using symmetry operations. The identification of type-II Weyl points for this class of compounds has not been documented in previous studies. The details of coordinates of Weyl points can be found in supplementary material (Sec. IIC) \cite{supp}. Fig. \ref{fig:nsoc} (d) shows type-II Weyl crossing in PdAsS at about 0.49 eV (E$_{WP}$) below E$_F$, while in other two materials it appears at about $\sim$-0.55 eV as shown in supplementary Sec II. A \cite{supp}.
 
 Ideally, energy dispersion of spin-1 excitation feature two linearly dispersing bands with slopes symmetric about the point of touching and a middle flat band but a little deviation form ideal characteristic dispersion is expected in real materials. Consequentially it is seen that there is a significant reduction in region resembling ideal dispersion of spin-1 excitation, decreasing from about 70 meV in PdAsS to merely 10 meV in PdBiTe. It is also observed that the middle flat band deviates from it ideally flat character and gains curvature from PdAsS to PdBiTe. Further, in all the materials considered here we see a slight variation in slopes of bands lying above and below E$_{S1}$; depicted in xy-plane for PdAsS in Fig. \ref{fig:ndisp} (a). At the point R, the double Weyl cones remain fairly symmetrical about the point in low energy region $\sim$20 meV about E$_{DWP}$ $-$a feature that remains almost identical in all the three materials. Fig. \ref{fig:ndisp} (b) displays dispersion in xy-plane around R- point. Similar three dimensional dispersions for PdSbSe and PdBiTe at $\Gamma$- and R-point are shown in Sec. IIA (Fig. S2) of \cite{supp}. Further, to establish type-II nature of the Weyl points, we plot energy dispersion in xy-plane around Weyl point as shown in Fig. \ref{fig:ndisp} (c) for PdAsS, where tilted Weyl cones with bands disperse linearly away from Weyl point. Bands lying above and below E$_{WP}$ cross the isoenergy surface (with E$_{WP}$ scaled to zero in Fig. \ref{fig:ndisp} (c)) . In, PdSbSe and PdBiTe the energy dispersion around Weyl point (see Fig. S3 \cite{supp}) features highly tilted Weyl cones with relatively shallow slopes as compared to PdAsS. 

Motivated by the observed deviation from characteristic dispersion of spin-1 excitations (mainly in middle flat band) across the materials, we analyze the orbital character contributions (OCC) to the bands forming multifold nodes to study the underlying origin of this behavior. We plot OCC for all the three materials in high symmetric direction $X-M-\Gamma-R-X$. Fig. \ref{fig:nocc} shows OCC plots for PdAsS, while OCC plots for PdSbSe and PdBiTe are in Sec. IIA (Fig. S4) \cite{supp}. Bands 1-3, which give rise to spin-1 excitation, are primarily dominated by Pd-\textit{4d} character with a relatively small contribution from S-\textit{3p} states; approximately one-fifth of Pd-\textit{4d} contribution as shown in Fig. \ref{fig:nocc} for PdAsS. Further with increasing energies of bands from 1 to 3 contribution from As-\textit{4p} also increases from almost zero in band 1 to about 0.1 in band 3; see Fig. \ref{fig:nocc}. A weak to moderate hybridization is seen all three bands between \textit{d}- and \textit{p}-orbitals. PdAsS and PdSbSe show almost identical OCC feature as shown Fig. \ref{fig:nocc} and Fig. S3 (a) \cite{supp} . Whereas a substantial hybridization is observed in PdBiTe between Pd-\textit{4d}, Bi-\textit{6p}, and Te-\textit{4p} at $\Gamma$-point in all three bands (Fig. S4 (b) \cite{supp}). Generally, it is seen that flat bands are formed where the concerned band has one pure character i.e., atomic-like state and has a constant value of character in a region. This is observed in case of PdAsS and PdSbSe; where band 2 is dominated by \textit{d}-character. However, in PdSbSe the magnitude of this character remains constant over a considerably small region around $\Gamma$ compared to PdAsS. On the other hand, higher degree of hybridization between the orbitals in PdBiTe does not conform to the argument above. As a result it is seen that middle band in PdBiTe remains flat in a region extremely close to $\Gamma$-point and gives an impression of parabolic band in low energy scale. It can thus be concluded here that increasing hybridization adds a non-linear (almost quadratic) dispersion to middle flat band in spin-1 excitation in real materials. Meanwhile, at point-R a little dip in Pd-\textit{4d} character can be observed in bands 1-2 with a simultaneous peak in As-\textit{4p} character and a similar opposite trend can also be observed in bands 3-4 in both PdAsS and PdSbSe. On the other hand at R-point in PdBiTe, double-Weyl node has Pd-\textit{4d} and Bi-\textit{6p} as major contributors. 
    
  \begin{figure}[t]
  	\centering
  	\includegraphics[width=0.498\textwidth]{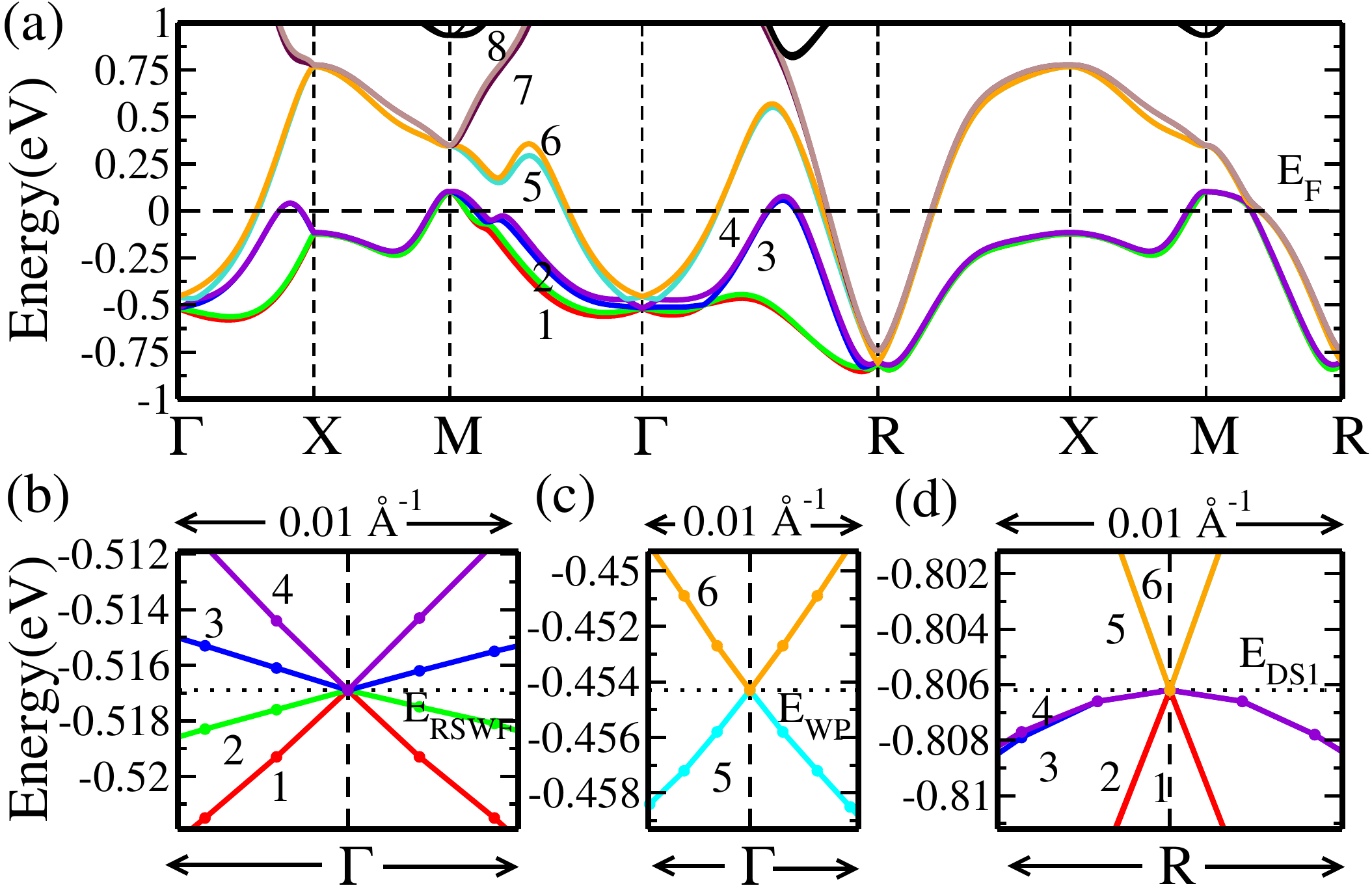}
  	\caption{Electronic band structures with spin-orbit coupling (SOC). (a) Full dispersion of PdAsS. (b) Zoom-in at $\Gamma$-point showing Rarita-Schwinger–Weyl fermion (RSWF) and type-I WP. (c) Zoom-in showing type-II Weyl points along $\Gamma$-R and $\Gamma-M$. (d) Zoomed view of double spin-1 excitation at R-point. }
  	\label{fig:WS}
  \end{figure}  
    
 In order to simulate the SOC effects in the materials on multifold nodes discussed above, the full band structure was calculated by treating SOC as a perturbation. Fig. \ref{fig:WS}(a) shows full band structure of PdAsS with SOC and E$_F$ set to zero. The electronic structure for PdSbSe and PdBiTe are presented in Sec. IIIA of supplementary material \cite{supp}. Note that, the inclusion of SOC in the materials induces a momentum-dependent spin splitting of bands away from time-reversal invariant momenta (TRIM), \textit{viz.} $\Gamma$, R, M, and X in all the three materials. At the same time bands do not split on the BZ boundaries ($M-X$ and $R-M$), and remain pairwise degenerate. This pairwise degeneracy is protected by screw symmetry \cite{bradlyn2016beyond}. A number of symmetry-enforced multifold nodes can be observed at $\Gamma$-, R- and M-point. At $\Gamma$, the six states originating from the threefold degeneracy without SOC split in two manifolds featuring a fourfold (spin-3/2, RSWF) and twofold (spin-1/2, Weyl) nodes. The RSWP (bands 1-4, Fig. \ref{fig:WS}(b)) and Weyl point ( bands 5-6, Fig. \ref{fig:WS}(c)) appear at energies $\sim -0.52$ and $\sim -0.45$ eV in PdAsS, respectively. In PdSbSe, the two manifolds can be found at energies $\sim$0.86 and $\sim$0.80 eV below E$_F$. Only in PdBiTe it is seen that the Weyl point ($\sim -0.89$ eV) lies lower in energy than RSWP ($\sim -0.71$ eV). At the same time, a six- and twofold degeneracy result from original fourfold degenerate state at R-point. While the sixfold node is known to possess topological character (characterized as double spin-1 excitation), the twofold degeneracy at R-point is a Kramer's doublet \cite{notes1}. The sixfold degeneracy in PdAsS appears at $\sim$0.8 eV below E$_F$ (bands 1-6, Fig. \ref{fig:WS} (d)). Meanwhile, the sixfold node in other two materials are located between (-0.7, -0.9 eV). In addition to the multifold nodes discussed above, a charge-2 fourfold node is also observed at M-point in al the three materials, lying between 100- 230 meV above E$_F$. Table \ref{tabb} lists respective energy difference between the manifolds at high symmtery points $\Gamma$ and R along with chirality of individual bands. In addition to the well known multifold nodes, our exploratory calculations using \textit{PY-Nodes} code also reveal a number of Weyl points at general positions in these materials which have not been reported in any of the previous studies \cite{PhysRevB.99.241104,PhysRevMaterials.9.L031201} in the same class of materials. Moreover, we also find Weyl points on high symmetry line $\Gamma-$R. A total of 12 type-II Weyl points are found at general momentum ($k_x,k_y,k_z$)$\frac{2\pi}{a}$ in all three materials with all of them lying at same energy. In contrast to this, the number of Weyl points along $\Gamma-R$ varies between the materials. While PdAsS and PdSbSe feature eight Weyl points on $\Gamma-R$ line, no such points are found in PdBiTe. We identify two distinct sets of Weyl points on $\Gamma-R$, namely, WP1 (($k_0,k_0,k_0$)$\frac{2\pi}{a}$) and WP2 (($-k_0,-k_0,-k_0$)$\frac{2\pi}{a}$) each consisting of four Weyl points in PdAsS and PdSbSe. All of these points lie at same energy in both sets and materials. Despite same structural symmetry across the materials, difference in number of Weyl points suggests that these degeneracies may be accidental in nature. To verify the topological nature of these points we have also calculated the associated chirality. The details of coordinates of all Weyl points are given in Sec. IIIC of supplementary material \cite{supp}. 
    
  \begin{table}[h]
 	\caption{\label{tabb}Relative energy splittings (in meV) between the manifolds at the high-symmetry points are listed, together with the corresponding chirality of the bands forming multifold points. Band indices are mentioned on corresponding band structures, and WP (Weyl point) }
 	
 	\begin{ruledtabular}
 		\begin{tabular}{llll}
 			\textrm{{Material}}&
 			\textrm{{PdAsS}}&
 			\textrm{{PdSbSe}}&
 			\textrm{{PdBiTe}}\\
 			\colrule
 			&&&\\
 			$\Delta E_\Gamma$&62.51&56.06&176.88\\
 			$\Delta E_R$&63.22&5.16&252.24\\
 			C (RSWP at $\Gamma$)&-3, -1,1,3&-3, -1,1,3&3,1,-1,-3\\
 			bands& 1-4&  1-4&  3-6\\
 			C (Type-I WP at $\Gamma$)&-1,1&-1,1&1,-1\\
 			 bands& 5-6& 5-6&1-2\\  		 
 			C(double spin-1 at R)&4,0,-4&4,0,-4&-4,0,4\\
 			bands (pairwise)&1-6 &3-8&3-8\\
 			
 		\end{tabular}
 	\end{ruledtabular}
 \end{table}
  
Although electronic structure provides an initial insight into a material's behaviour, it is the low-energy dispersion around point of interest that reveals true underlying toplogical features associated with them. The low-energy dispersion is plotted for all the materials around multifold as well as Weyl nodes, presented for PdAsS (Fig. \ref{fig:TOF}) in the main text and rest in supplementary (Sec. IIIB, Fig. S6 $\&$ S7) \cite{supp}. Notably, the dispersion around RSWP at $\Gamma$-point (shown for PdAsS in Fig. \ref{fig:TOF}(a)) remains almost identical across the materials, with little deviation from linearity in middle bands of RSWF in PdBiTe. This deviation may be attributed to strong hybridization at $\Gamma$ in PdBiTe as evident from its OCC plot (Fig. S8 (b) \cite{supp}). In contrast, at the point-R where sixfold degeneracy (double spin-1 excitation) is known to arise from the combined action of the little group symmetries and time-reversal symmetry a maximum deviation from idealized behavior is observed. In PdAsS (Fig. \ref{fig:TOF} (b)) and PdBiTe, due to hybridization the middle bands remain flat only in the immediate vicinity of the R-point before exhibiting almost parabolic dispersion as shown in their OCC plots. Conversely, the middle bands in PdSbSe disperse linearly with a positive slope. This behavior may be an outcome of rapidly changing orbital characters at R in PdSbSe (band 6-7, Fig. S8 (a) \cite{supp}). The discussion above has been supported with OCC plots for all bands forming multi fold nodes in the three materials; for PdAsS it is shown in Fig. \ref{fig:socc}. OCC plots for PdSbSe and PdBiTe are depicted in Sec. IIID of supplementary material \cite{supp}. It is also interesting to note that even with such deviation the topological character remains intact i.e. chirality of bands is well defined; refer to Table \ref{tabb}. Low-energy dispersion around R-point in PdSbSe and PdBiTe are shown in Fig. S6 of supplementary material \cite{supp}. Further, three dimensional dispersion of Weyl nodes also confirms their type-II nature as shown in Fig. \ref{fig:TOF} (c) and (d) for PdAsS. 

  \begin{figure}[t]
	\captionsetup[subfigure]{labelformat=empty}
	\centering
	\begin{subfigure}[b]{0.25\textwidth}            
		\includegraphics[width=\textwidth,height=3.5cm]{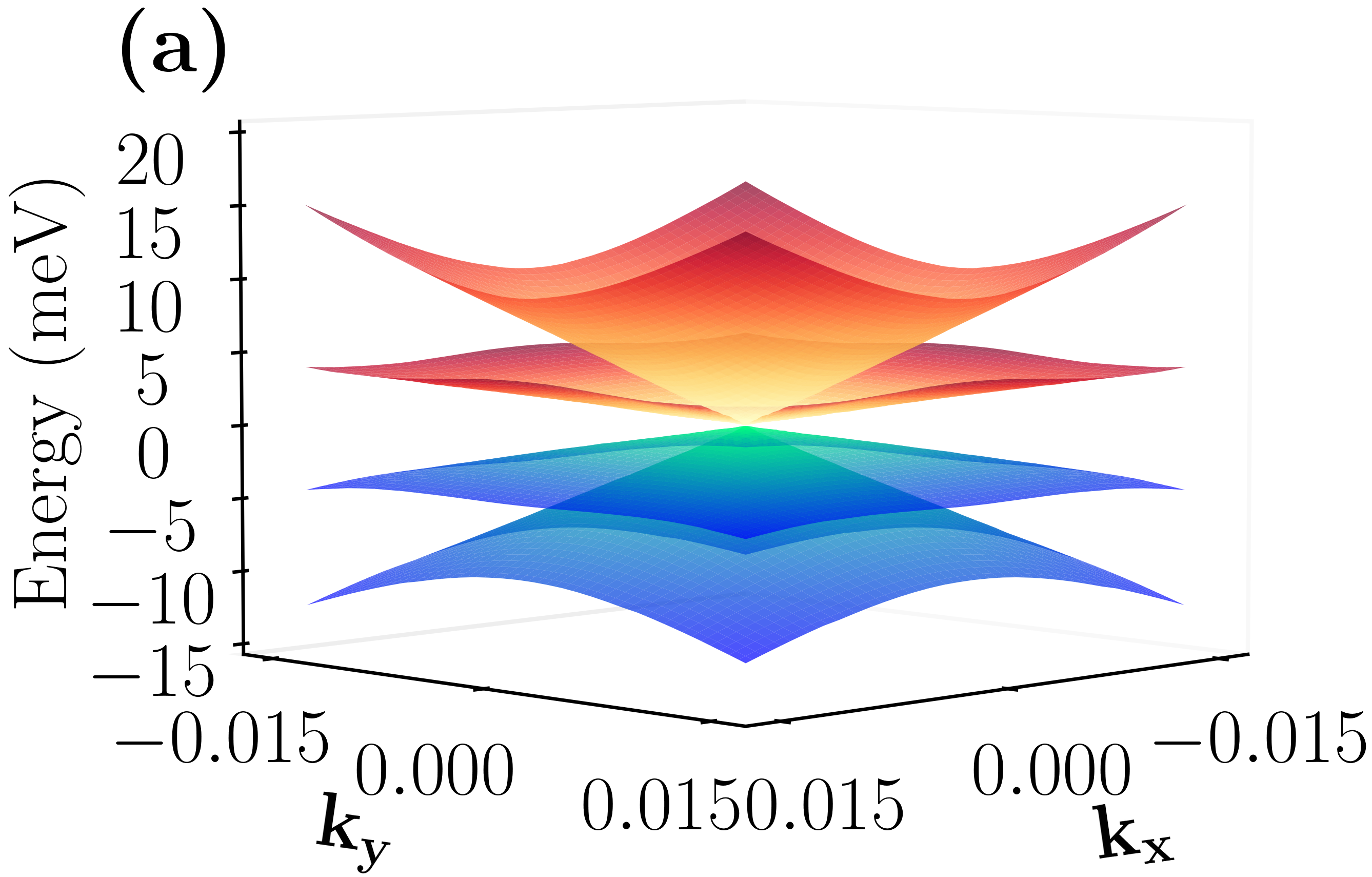}
		\caption{\hspace{-1em}}		
	\end{subfigure}%
	\begin{subfigure}[b]{0.25\textwidth}
		\centering
		\includegraphics[width=\textwidth,height=3.5cm]{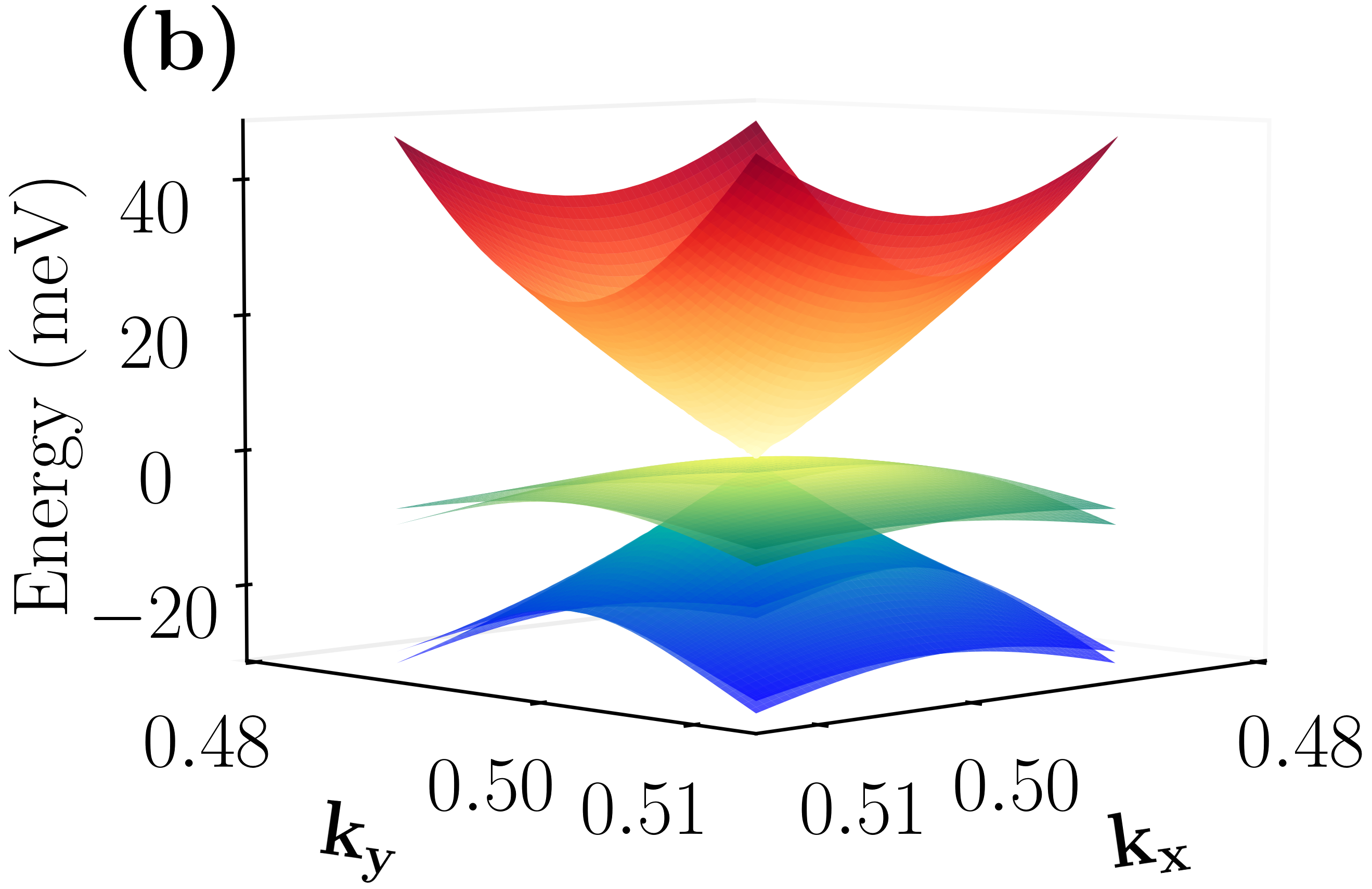}
		\caption{\hspace{-1em}}		
	\end{subfigure}
	\begin{subfigure}[b]{0.235\textwidth}
		\centering
		\includegraphics[width=\textwidth,height=3.5cm]{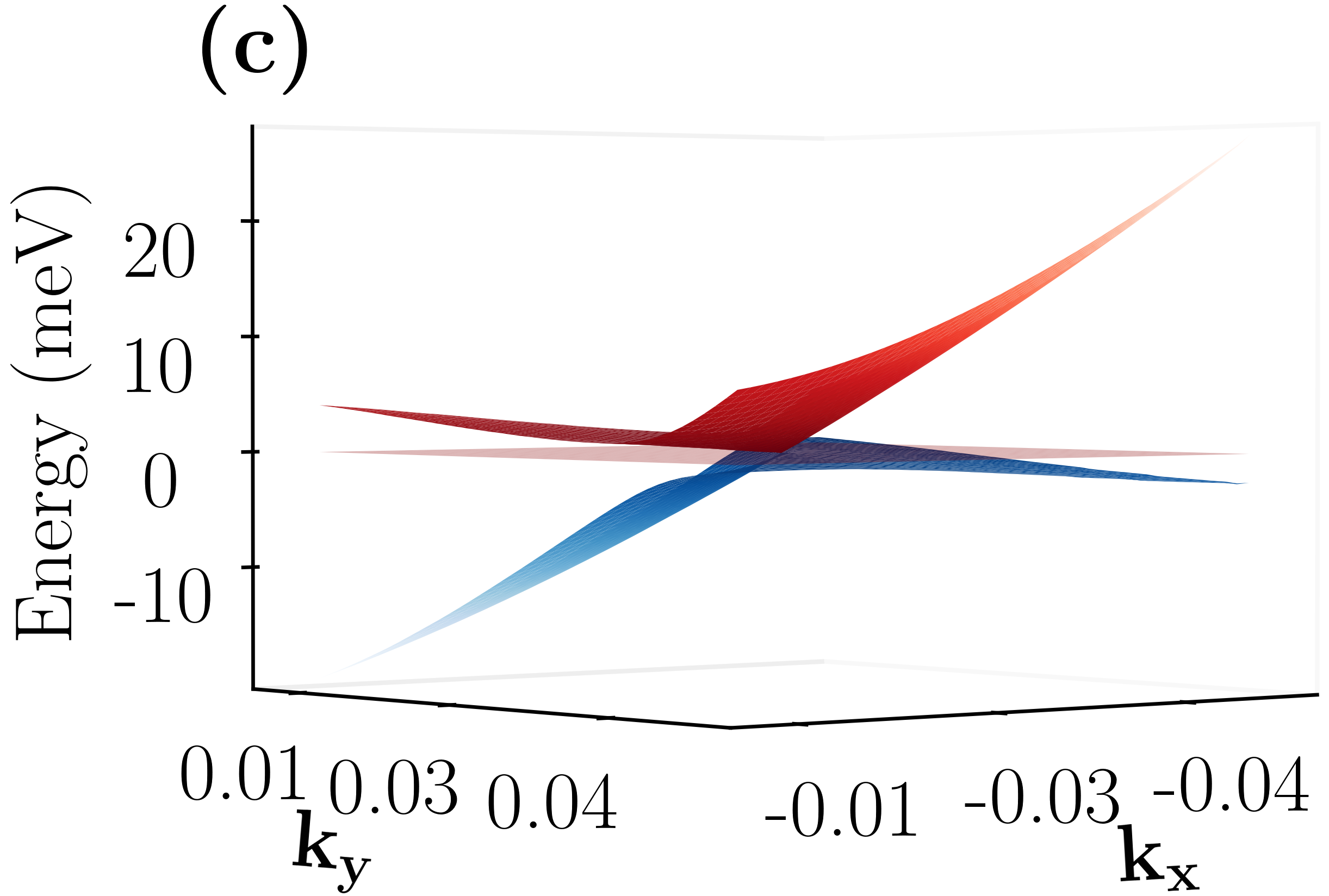}
		\caption{\hspace{-1em}}		
	\end{subfigure}	
	\begin{subfigure}[b]{0.24\textwidth}
		\centering
		\includegraphics[width=\textwidth,height=3.5cm]{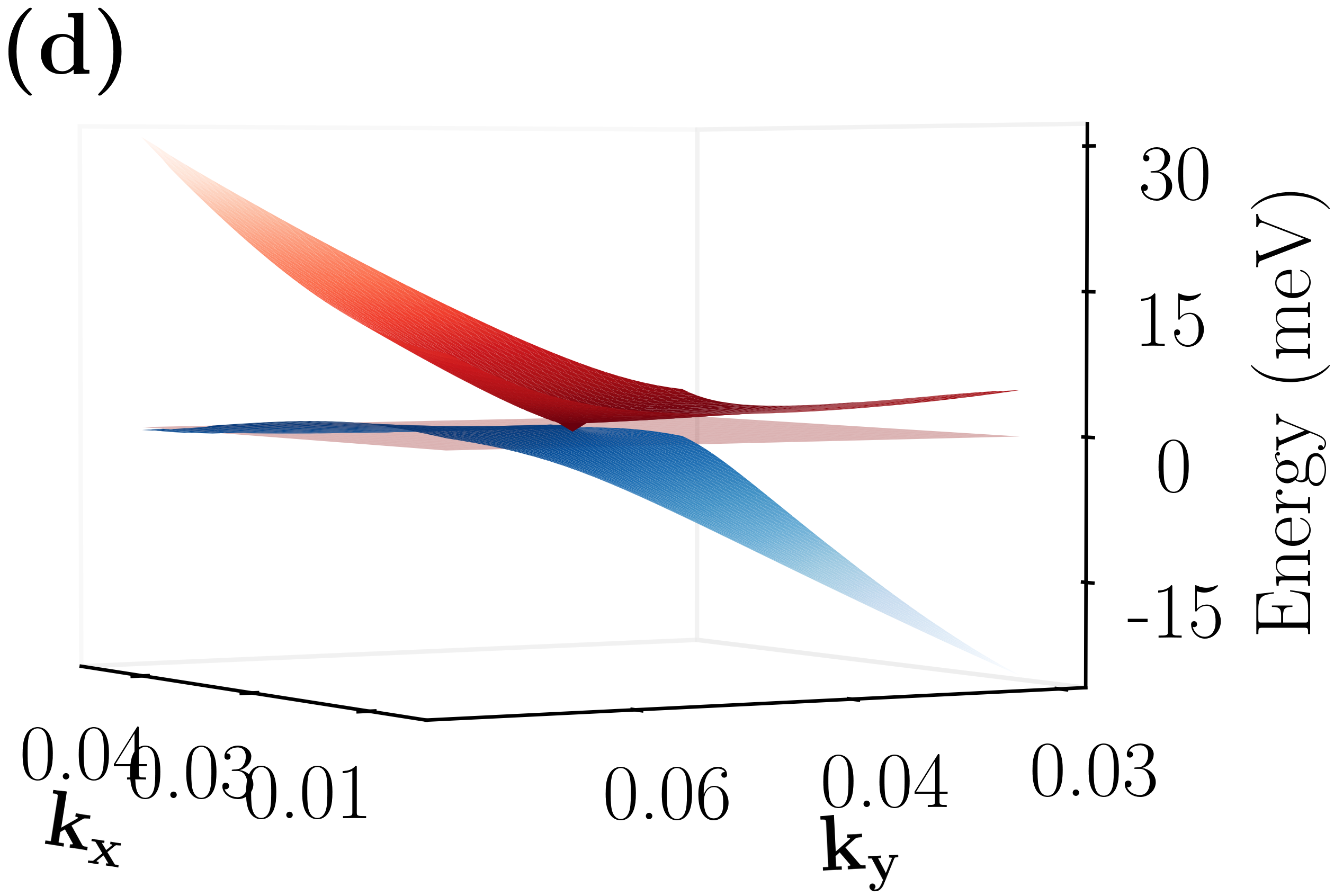}
		\caption{\hspace{-1em}}		
	\end{subfigure}
	\caption{The figure displays energy dispersion in xy-plane near the multifold nodes at $\Gamma$- and R- point in the BZ of PdAsS. (a) Rarita-Schwinger-Weyl excitation (spin-3/2 fermion); (b) Double spin-1 excitation; (c) and (d) Energy dispersion around Weyl point in xy-plane on $\Gamma-R$ line and general position in PdAsS, respectively. z-axis shows energy in meV. Note that, the energies of multifold and Weyl nodes are scaled to zero here. }\label{fig:TOF}
\end{figure}

\begin{figure}[t]
  	\centering
  	\includegraphics[width=0.48\textwidth]{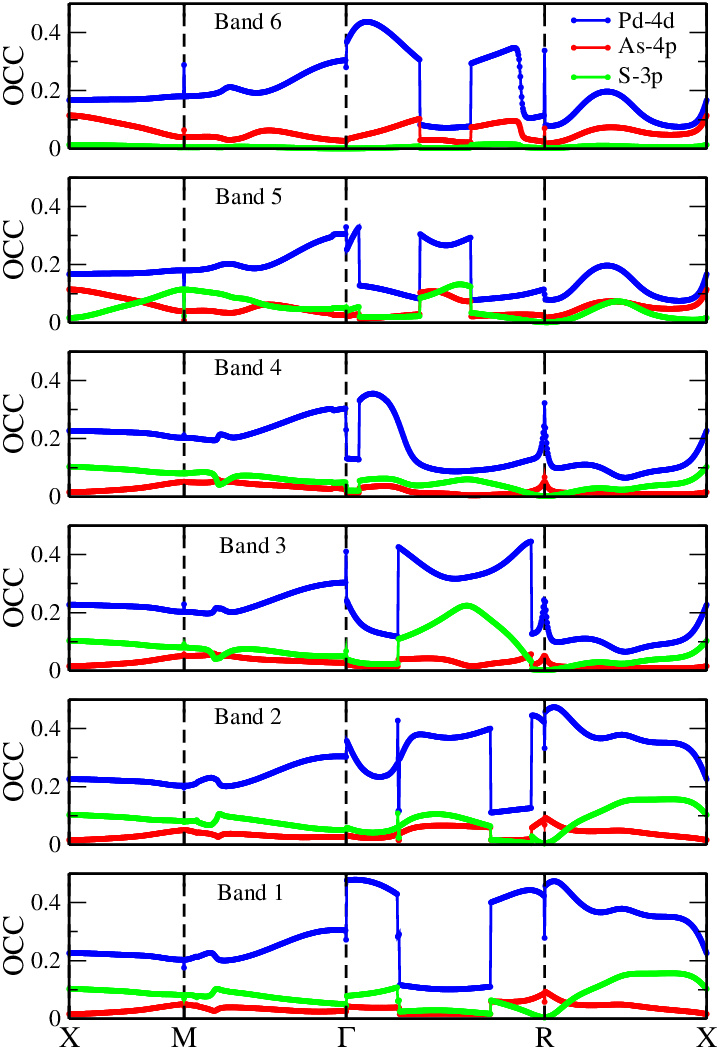}
  	\caption{Projected orbital contributions to the bands constituting multifold nodes in presence of SOC. Band 1-4 give RSWP at $\Gamma$ and Band 1-6 form double spin-1 excitation at R}
  	\label{fig:socc}
\end{figure}
 \begin{figure}[ht]
	\includegraphics[width=0.495\textwidth]{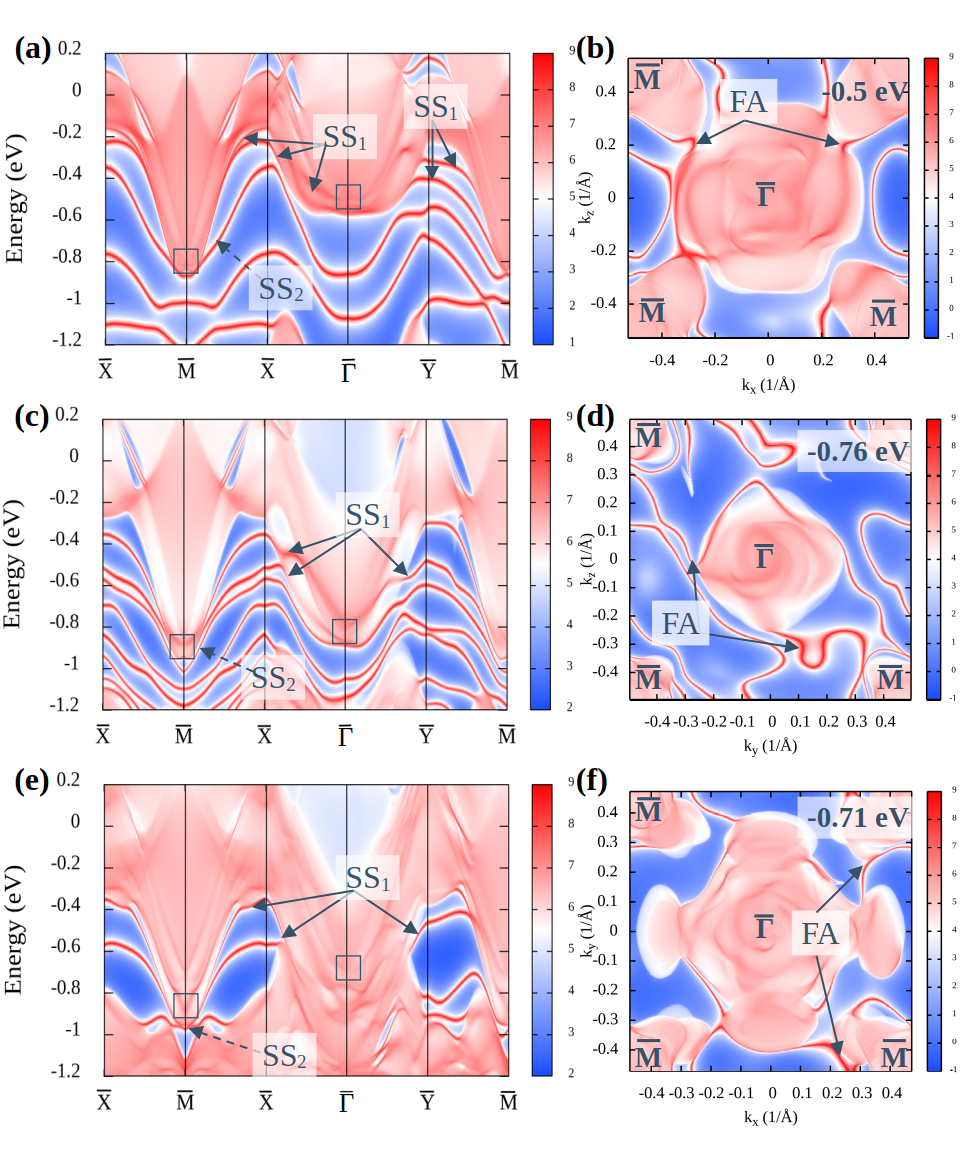}
	\caption{The surface states and Fermi arcs obtained (a) and (b) PdAsS on (010) surface; (c) and (d) PdSbSe on (100) surface; and (e) and (f) PdBiTe (001) surface. Non trivial surface states are marked by arrows on the spectra.}
	\label{fig:WT}
\end{figure}

\subsection{Surface States and Fermi arcs}
Topological excitations are fundamentally characterized by their distinctive low-energy dispersion (often linear) and topological invariants such as Chern numbers, topological charges, and chirality. Another fundamental signature often associated with non-trivial bulk topology is the emergence of protected surface states(SS) and Fermi arcs (FA) on one or more surfaces of the BZ. These SS are the direct manifestation of bulk-boundary correspondence in topological materials \cite{vanderbilt2018berry}. Accordingly, study of surface spectra and arcs is essential for understanding the topological nature and properties of a material. We thus present our SS and FA for all the three materials in Fig. \ref{fig:WT}. The surface spectra are presented in Fig. \ref{fig:WT}(a), (c) and (e) for PdAsS, PdSbSe, and PdBiTe, respectively. The bulk TRIM points $\Gamma$ and R are projected onto $\overline{\Gamma}$ and $\overline{\text{M}}$ , respectively, on the surface BZ. The projections of multifold nodes are marked by box on respective plot. From the surface spectra plots (Fig. \ref{fig:WT}(a), (c) and (e)) two distinct types of SS can be identified i.e., SS$_1$ and SS$_2$ in all three materials. SS$_1$ which are marked by solid arrows on the plot span between region around $\overline{\Gamma}$ and $\overline{\text{M}}$ in all three materials. Shadows of surface bands extending into the bulk bands can also be observed around $\overline{\Gamma}$ in all the three materials; most prominent in PdSbSe. On close observation of SS$_1$, it appears that it passes through projection of multifold nodes RSWF ($\overline{\Gamma}$) and double spin-1 excitation ($\overline{\text{M}}$) in PdAsS and PdSbSe. SS$_1$ also seems to be partially obscured around $\overline{\Gamma}$ by the projection of bulk bands on the respective surfaces in the materials. On the other hand, connection between $\overline{\Gamma}$ and $\overline{\text{M}}$ is hard to observe in PdBiTe where heavy projections of bulk bands seem to conceal surface states around $\overline{\Gamma}$ completely. It is interesting to note that SS$_1$ imply a connection between two different kind of topological excitations i.e., RSWF and double spin-1 excitation which carry same magnitude of monopole charge, 4. This connection establishes the non-trivial nature of these SS. Similar results have also been reported in previous studies of materials belonging to same space group \cite{PhysRevLett.119.206402, PhysRevLett.119.206401, rao2019observation, schroter2019chiral}. Further we also note that SS$_1$ in these materials are fairly dispersive, with a band width of $\sim$ 400 eV in PdAsS to $\sim$ 200 in PdSbSe and PdBiTe. Next, we consider other non-trivial surface states, SS$_2$, which connects two different $\overline{\text{M}}$ points on surface BZ i.e., $\overline{\text{M}}$ (0.5,0.5) and $\overline{\text{M}}$ (-0.5,0.5). SS$_2$ in PdAsS can be observed distinctly, whereas in PdSbSe SS$_2$ connects $\overline{\text{M}}$-points through the perimeter of bulk band projection around $\overline{\Gamma}$ (Fig. \ref{fig:WT}(c)). In comparison to the two materials, the SS$_2$ in PdBiTe are relatively faint and can only be observed in the vicinity of $\overline{\text{M}}$-points; Fig \ref{fig:WT}(e).

Fig. \ref{fig:WT}(b) shows isoenergy plot on (010) surface in PdAsS at energy -0.5 eV ($\sim$E$_{RSWF}$), at $\Gamma$. Although faint, FA can be observed emerging from region around $\overline{\text{M}}$ and merging into $\overline{\Gamma}$; marked by arrows on the plot. Similar FA are also observed in the case of PdBiTe as shown in Fig. \ref{fig:WT}(f) on (001) surface and at energy $\sim$ E$_{RSWF}$ (0.71 eV). Meanwhile, PdSbSe features longest and cleanest FA among these materials. Fig. \ref{fig:WT}(d) shows FA in PdSbSe on (100) surface at energy 0.76 eV below E$_F$. Appearance of such faint FA seem to arise from the nature of dispersion of bulk bands around multifold nodes. Here, we see that away from higher fold nodes bulk band split under SOC and ultimately disperse in the same direction in all three materials; Fig. \ref{fig:WS}(a) for PdAsS. As a result when these bulk bands are projected onto surfaces, they potentially fill in the surface gaps in the vicinity of monopoles tending to partially or fully mask visible trace of the non-trivial FA \cite{PhysRevB.99.241104,chang2017unconventional}; as also evident from Fig. \ref{fig:WT}(b) and (f). Thus, the resolution of FA on a given surface depends critically on the bulk band dispersion and does not depend solely upon the presence of monopole charges. Another plausible reason might be the dispersive nature of SS, which also makes it difficult to resolve FA on one particular energy. Further, the FA associated with the Weyl points cannot be resolved due to reasons mentioned above since these points lie very close to $\Gamma$-point.

 
\section{Conclusions}
Our DFT based study confirms a presence of rich variety of unconventional as well as known excitations i.e., double Weyl, Rarita-Schwinger-Weyl fermions, spin-1, double spin-1 and type-II Weyl in PdAsS, PdSbSe and PdBiTe. We identify several previously undocumented type-II Weyl points in these materials in absence and presence of SOC. Twelve new Weyl points are found at general momenta in these materials. Further, based on the analysis of low-energy dispersions of unconventional excitations and surface spectra, there are two main remarks. First, varying deviation among materials in low-energy dispersion of spin-1 and double spin-1 excitations from their characteristic behavior is obeserved. We find that orbital characters and hybridization play a crucial role. Both in case of spin-1 and double spin-1, we find that ideally flat middle band is parabolic in PdBiTe. Astoundingly, middle bands of double spin-1 excitation in PdSbSe are linearly dispersive. Second, while surface spectra shows several non-trivial surface states in these materials related Fermi arcs could only be resolved in PdSbSe. We find that bulk dispersion plays an important role in observability of Fermi arcs on a surface and the presence of topological charges does not necessarily guarantee observable Fermi arcs.

For future scope, due to inherent structural chirality of these materials a unique response to circularly polarized light is expected. Moreover, studying features of dispersion around unconventional excitation could prove to be essential in harnessing full potential of topological nodes and tuning band features using band structure engineering. 

\bibliography{main}
\bibliographystyle{apsrev4-2}

\clearpage
\onecolumngrid

	\large \centering \textbf{Supplementary Material for} \enquote{Unconventional excitations and orbital-driven low-energy dispersions in chiral topological semimetals PdAsS, PdSbSe, and PdBiTe: a \textit{first-principles} study}

\section*{I.\hspace{0.3cm}Lattice Parameters}

\begin{table*}[h]\label{st1}
	\caption*{Table ST1:
		\small{The optimized lattice parameters, atomic positions and R$_{MT}$ used for ground state calculations. }}
	\begin{ruledtabular}
\begin{tabular}{lcccc}
	\textrm{{Material}}&
	\textrm{{Lattice Parameters}}&
	\textrm{{Experimental values}}&
	\textrm{{Atomic positions }}& 
	\textrm{{R$_{MT}$} (a.u.)}\\

	\colrule
	&   &   &  & \\
	PdAsS    & $a = b = c = 5.9445 $ \AA  & $a = b = c = 5.9476 $ \AA ~ \cite{FOECKER200169} & \textit{Pd}:  0.99784259 0.99784259 0.99784259  & \textit{As}:  2.14\\
	
	&  $\alpha=\beta=\gamma=90\degree$& &	\textit{As}:   0.61366515 0.61366515 0.61366515 &\textit{Pd}: 2.29 \\
	
	& & & \textit{S}:  0.39200367 0.99784259 0.39200367 &\textit{Pd}: 1.93 \\

	&   & &   &  \\
	PdSbSe    & $a = b = c = 6.3242 $ \AA  & $a = b = c = 6.3229$ \AA ~  \cite{childs1973crystal}&\textit{Pd}:   0.99386377 0.99386377 0.99386377& \textit{Pd}:  2.49\\
	
	&  $\alpha=\beta=\gamma=90\degree$& &\textit{Sb}:  0.62638966 0.62638966 0.62638966 &\textit{Sb}: 2.49 \\
	
	& & & \textit{Se}:  0.38205667 0.38205667 0.38205667&\textit{Se}: 2.38 \\
	& &  &   &  \\
	PdBiTe    & $a = b = c = 6.6495 $ \AA  & $a = b = c = 6.6420 $ \AA ~ \cite{FOECKER200169} & \textit{Pd}:    0.00500935 0.00500935 0.00500935& \textit{Pd}:  2.5\\
	
	&  $\alpha=\beta=\gamma=90\degree$& &\textit{Bi}:  0.36815886 0.36815886 0.36815886 &\textit{Bi}: 2.5 \\
	
	& & & \textit{Te}:  0.62713249 0.62713249 0.62713249 &\textit{Te}: 2.5 \\

\end{tabular}
\end{ruledtabular}
\end{table*}
\section*{II.\hspace{0.3cm}without SOC}
\subsection{Electronic structures}

\begin{figure*}[h]
\includegraphics[width=0.48\linewidth]{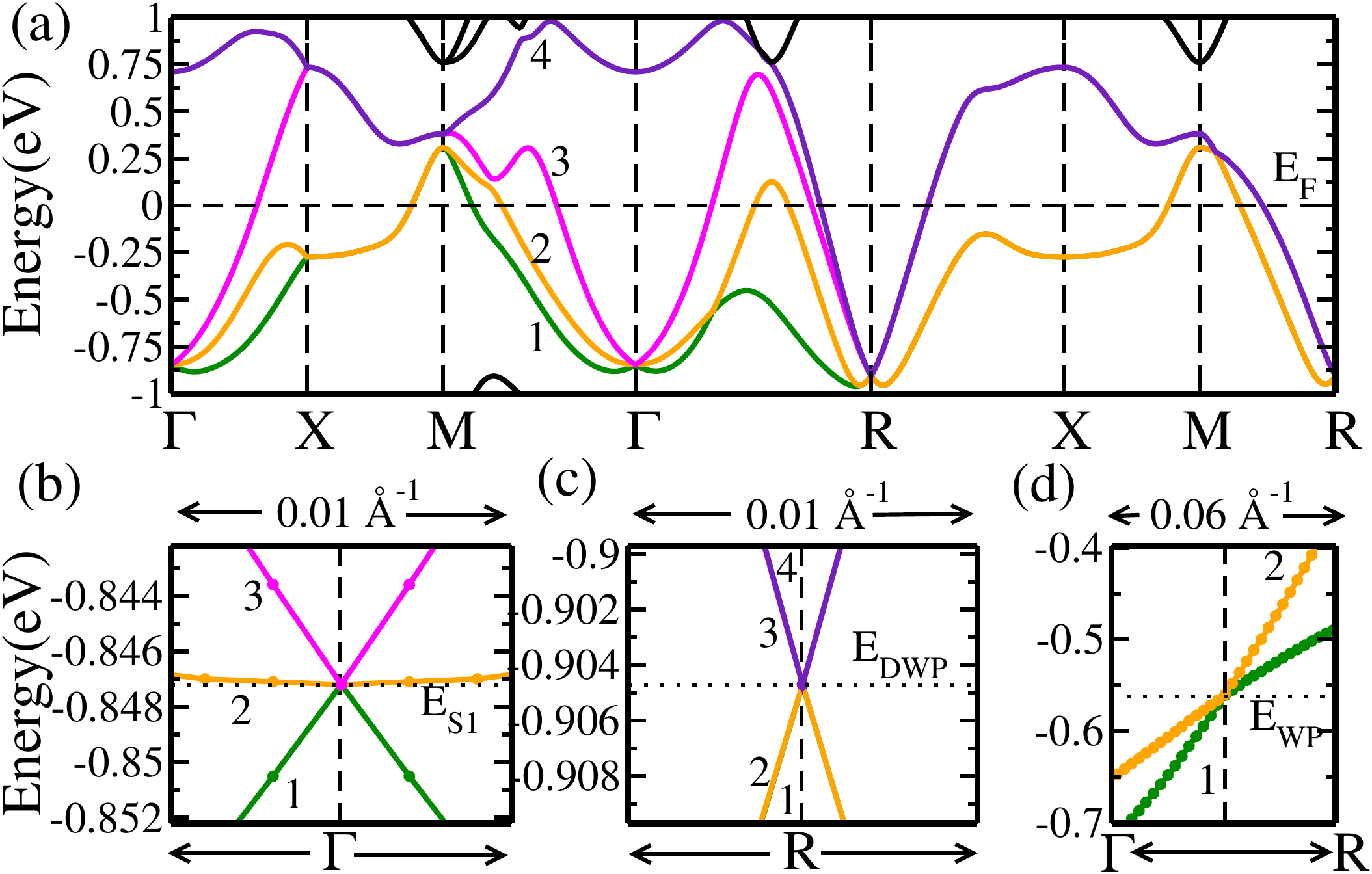}
\includegraphics[width=0.48\linewidth]{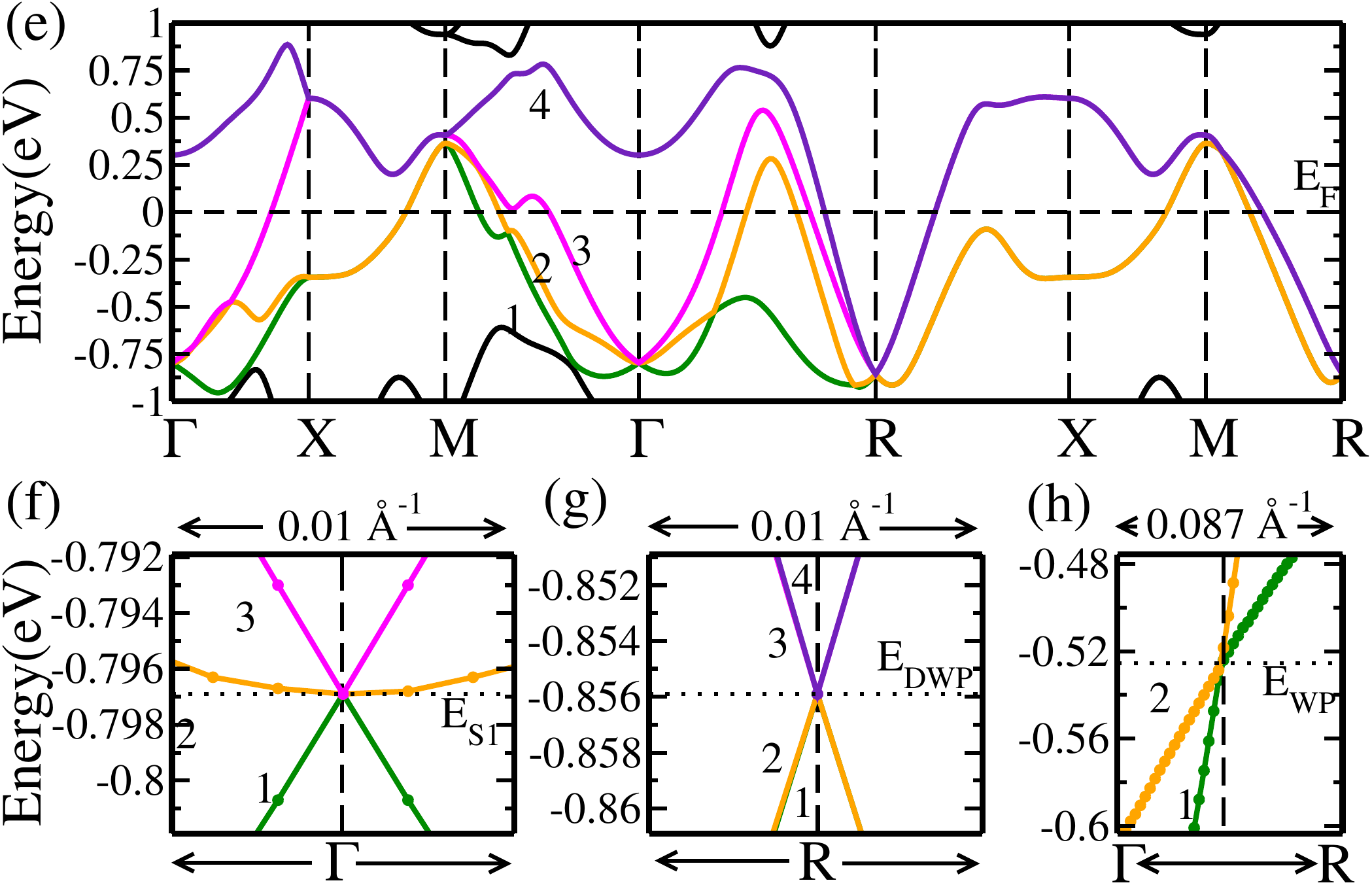}
\caption*{Fig. S1: Electronic structure in absence of spin-orbit coupling (SOC) in direction $\Gamma-X-M-\Gamma-R-X-M-R$ of (a) PdSbSe and (e) PdBiTe . (b) and (f) show a zoomed in view of dispersion around $\Gamma$-point featuring a spin-1 excitation; (c) and (g)  show dispersion around R-point showing double a Weyl point (Charge-2 four-fold fermion); (d) and  (h) show a type -II Weyl point (WP) along $\Gamma$-R}
\end{figure*}
\newpage

\subsection{Three dimensional energy dispersion}
\begin{figure*}[h]
\includegraphics[width=0.23\linewidth]{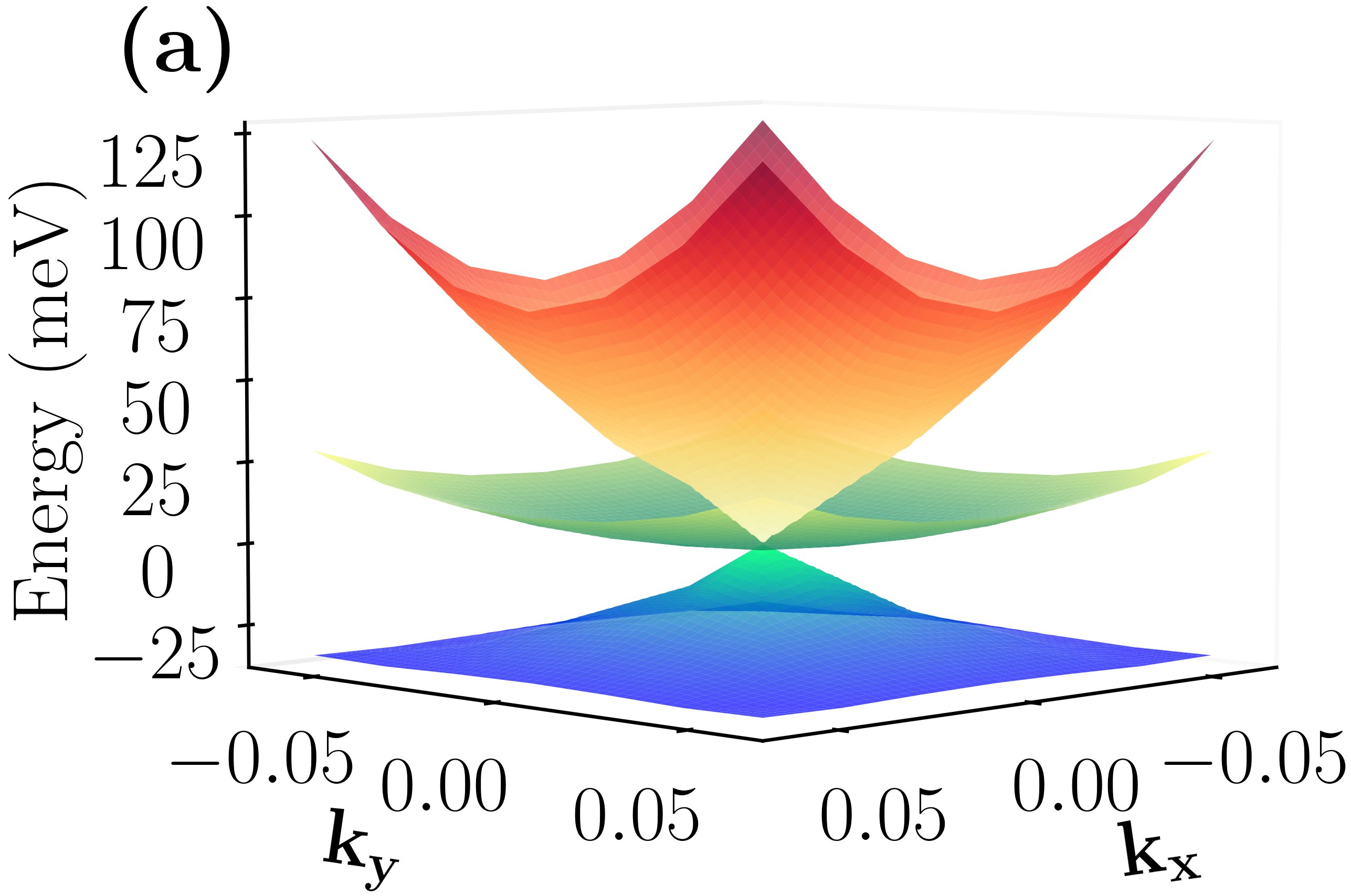}
\includegraphics[width=0.23\linewidth]{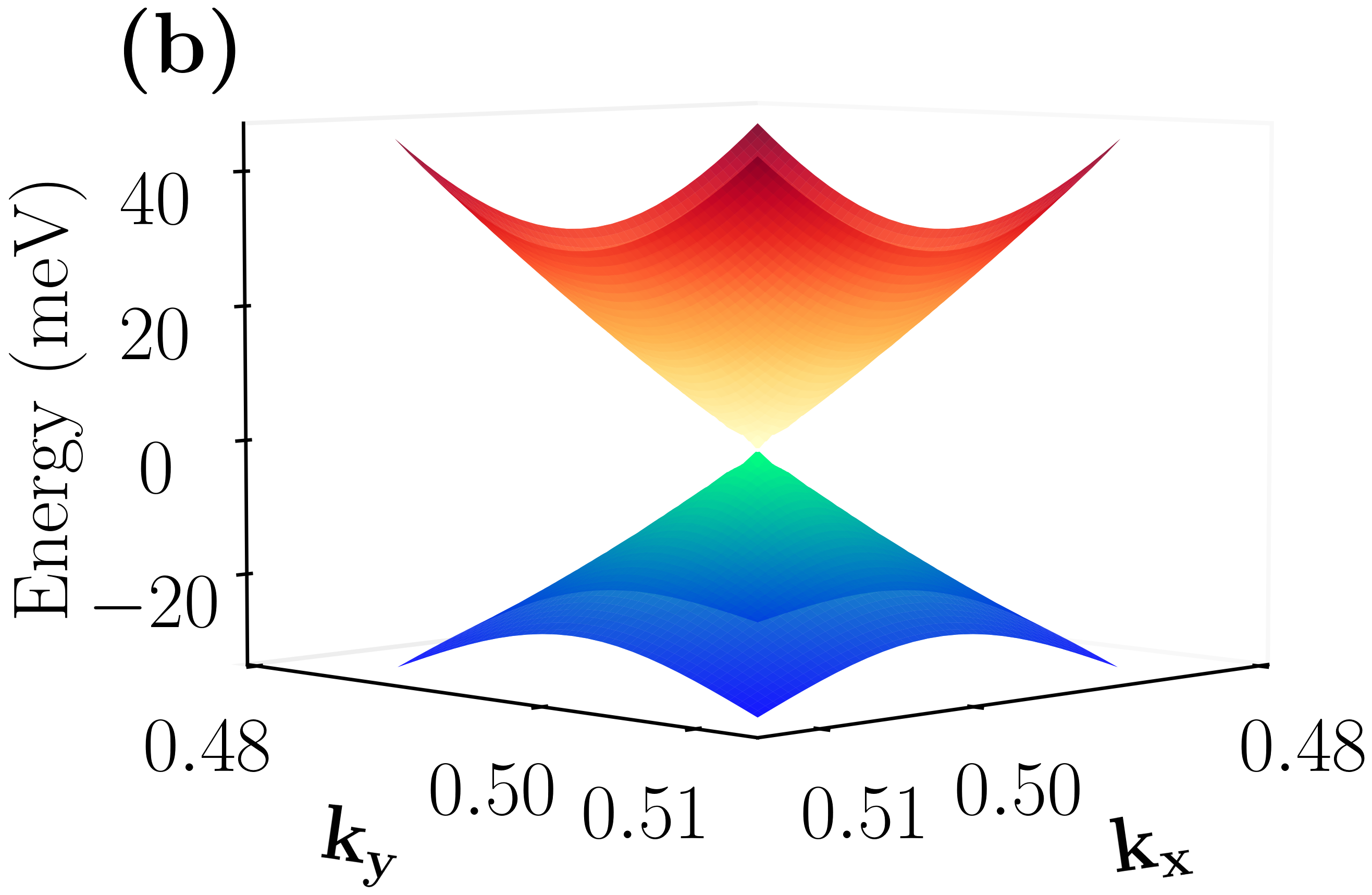}
\includegraphics[width=0.23\linewidth]{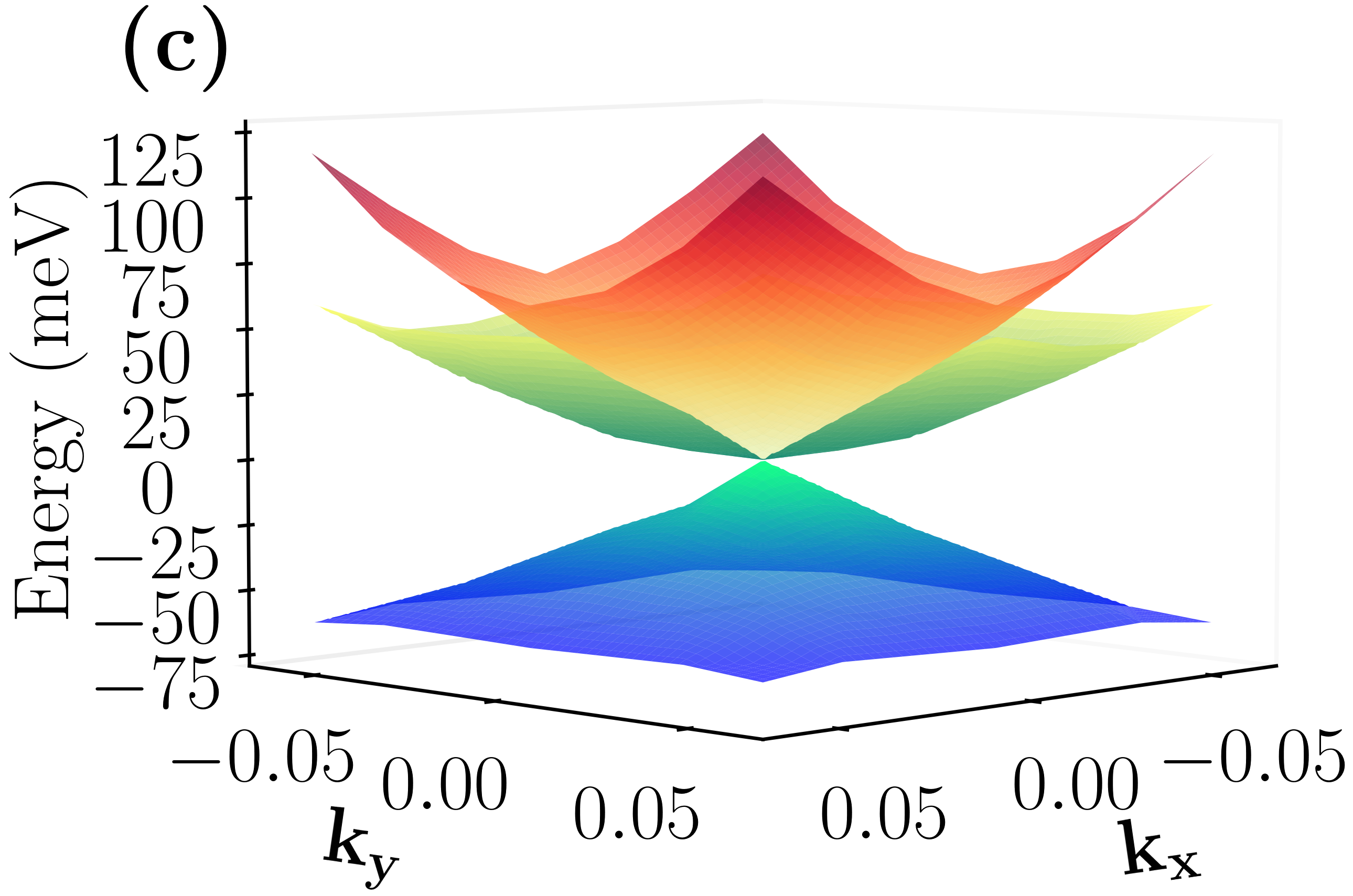}
\includegraphics[width=0.23\linewidth]{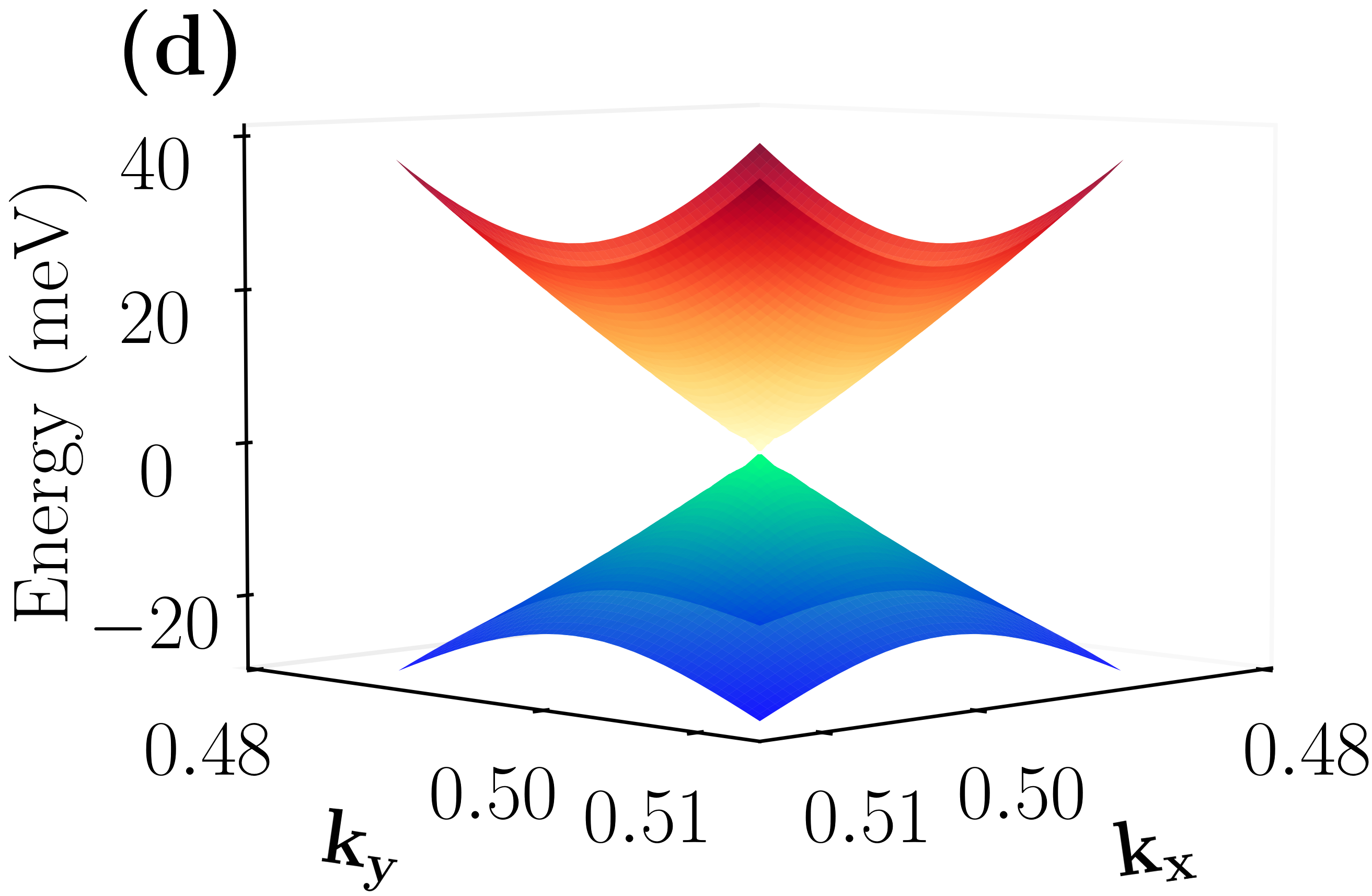}
\caption*{Fig. S2: Three-dimensional dispersion around high symmetric point $\Gamma$and R with out SOC. Spin-1 excitation at $\Gamma$ in (a) PdSbSe and (c) PdBiTe. A double Weyl point (Charge-2 four-fold fermion) at R-point in (b)PdSbSe and (d) PdBiTe.}
\end{figure*}

\begin{figure*}[h]

\includegraphics[width=0.32\linewidth,height=4cm]{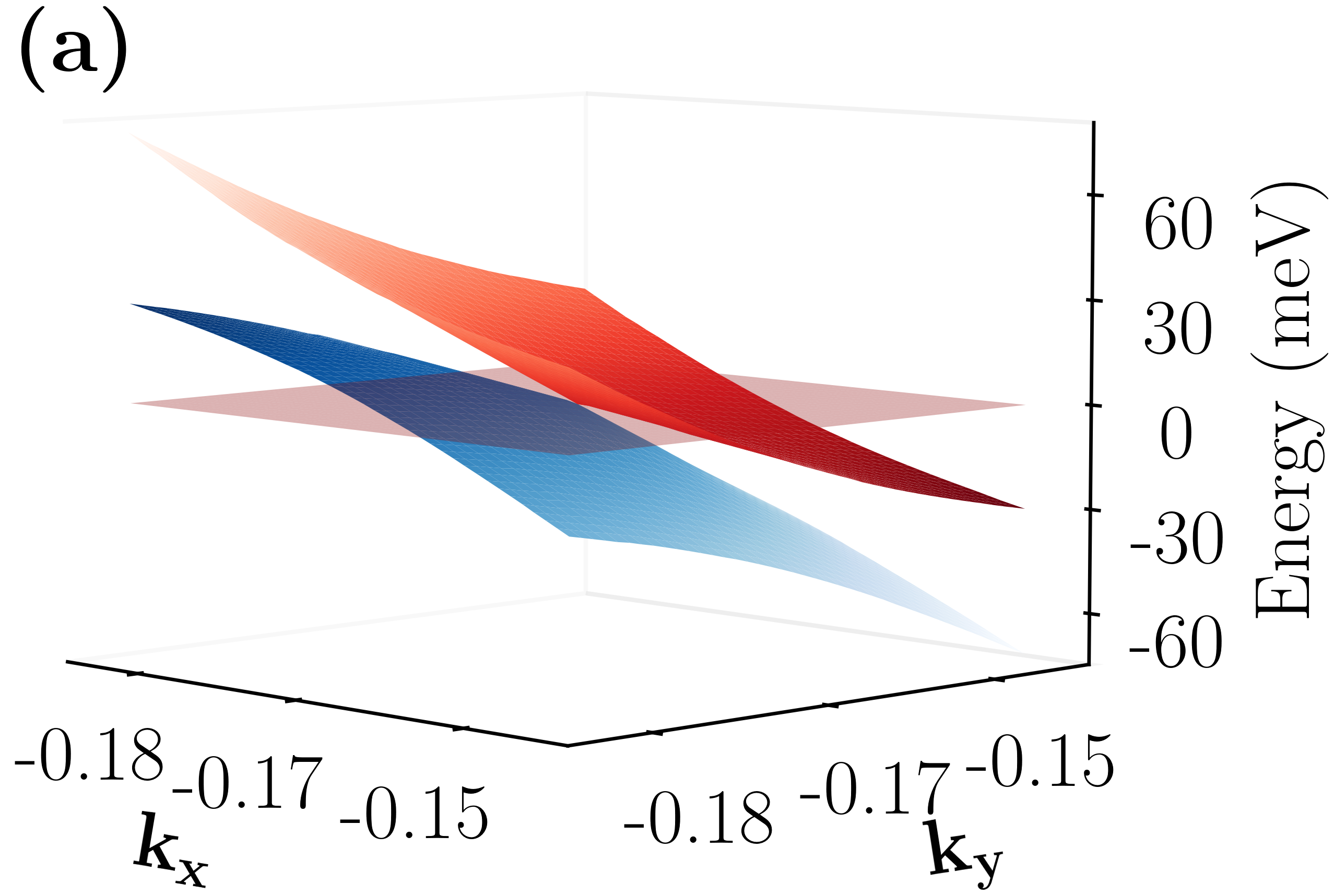}
\hspace{2.5cm}
\includegraphics[width=0.32\linewidth,height=4cm]{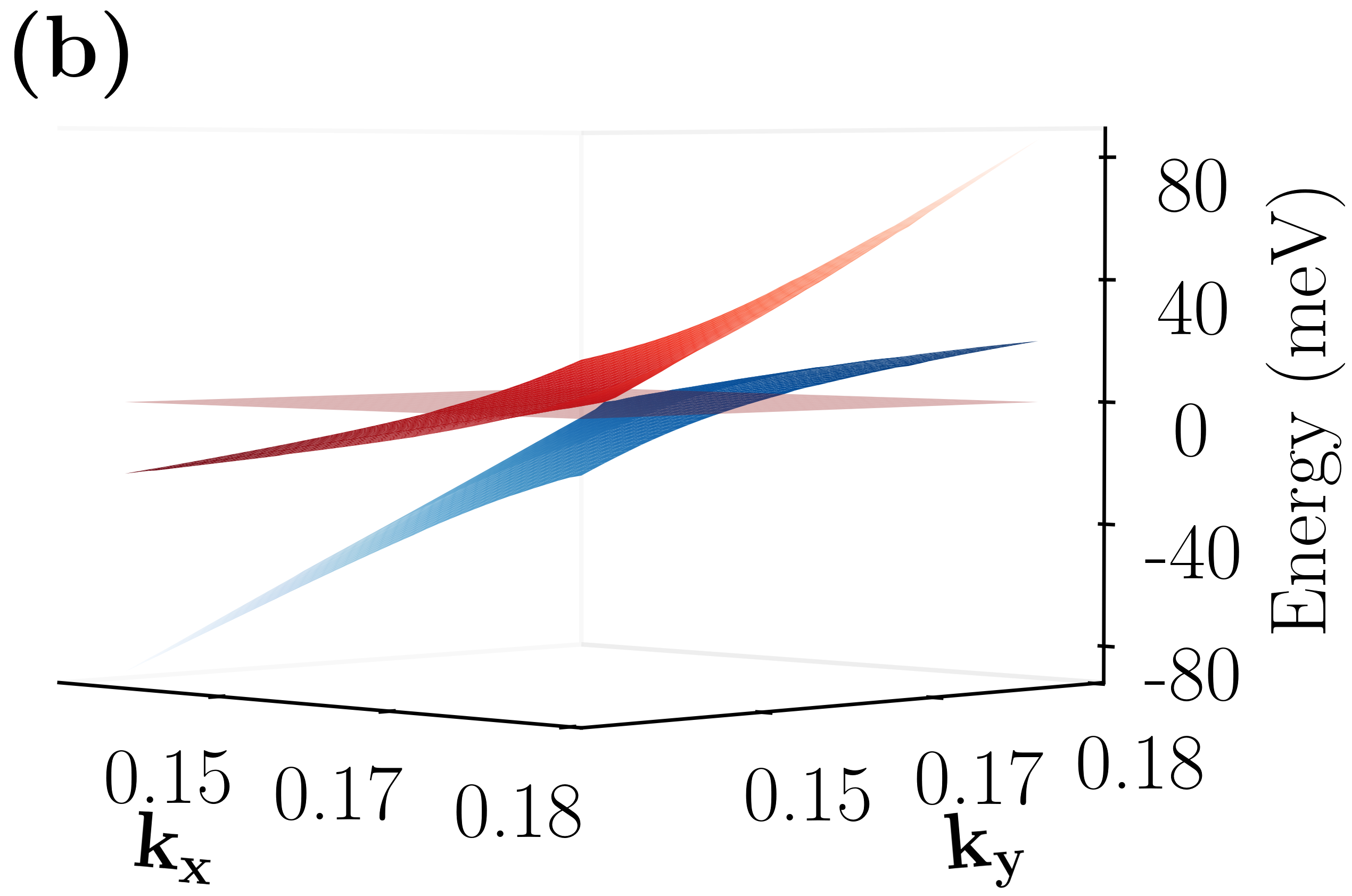}

\caption*{Fig. S3: Three dimensional energy dispersion around the obtained Weyl points. The Weyl point energy is set to zero and band energies are scaled accordingly. The grey plane represents the zero energy. 3-D dispersions are as follows : (a)  PdSbSe at point(-0.164, -0.164, -0.164)$\frac{2\pi}{a}$; (b) PdBiTe at point (0.159, 0.159, 0.159)$\frac{2\pi}{a}$ . Two linearly dispersing bands away from the Weyl points are seen and both bands cross the zero energy surface revealing type-II nature of Weyl points.}
\end{figure*}
\subsection{Weyl point coordinates}
\begin{table}[h]
\caption*{\label{tab:tab1}Table ST2:  The coordinates of Type-II WP in PdAsS, PdBiTe and PdSbSe}
\begin{ruledtabular}

\begin{tabular}{lcc}
	\textrm{Material}&
	\textrm{{WP coordinates along $\Gamma$-R}}&
	\textrm{{Energy in eV, Fermi energy $E_F$=0}} \\ 
	
	&($k_x,k_y,k_z$)$\dfrac{2\pi}{a}$&\\
	\colrule
	&&\\
	PdAsS  & ( $\pm$0.130, $\pm$0.130, $\pm$0.130)  & -0.488755  \\

	PdSbSe   & ($\pm$0.164, $\pm$0.164, $\pm$0.164) &  -0.5614720 \\
	
	PdBiTe  &   ($\pm$0.159, $\pm$0.159, $\pm$0.159)  & -0.5270437  \\
	
\end{tabular}
\end{ruledtabular}
\end{table} 
\newpage
\subsection{Orbital character plots}
\begin{figure*}[h] 
\includegraphics[width=0.48\linewidth]{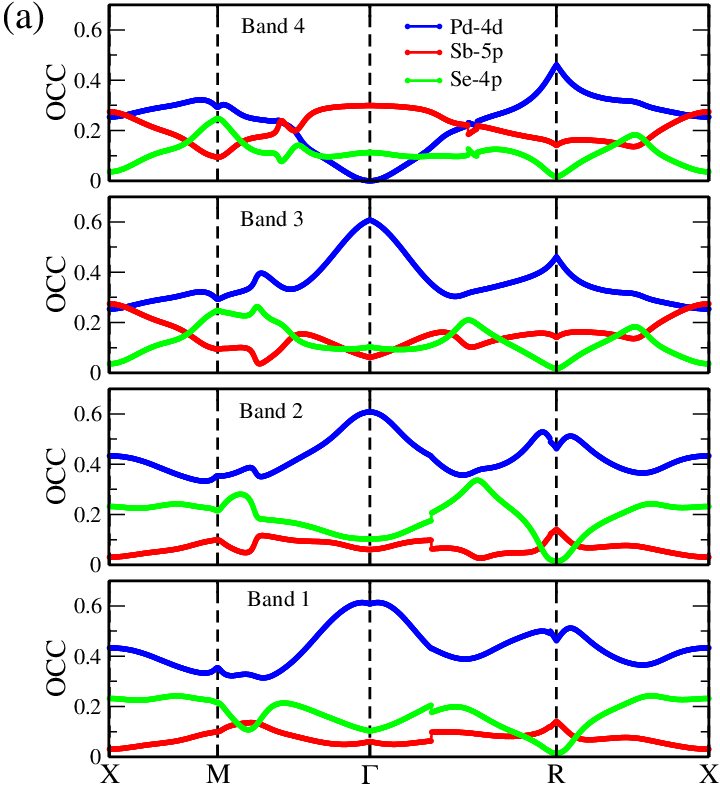}
\includegraphics[width=0.48\linewidth]{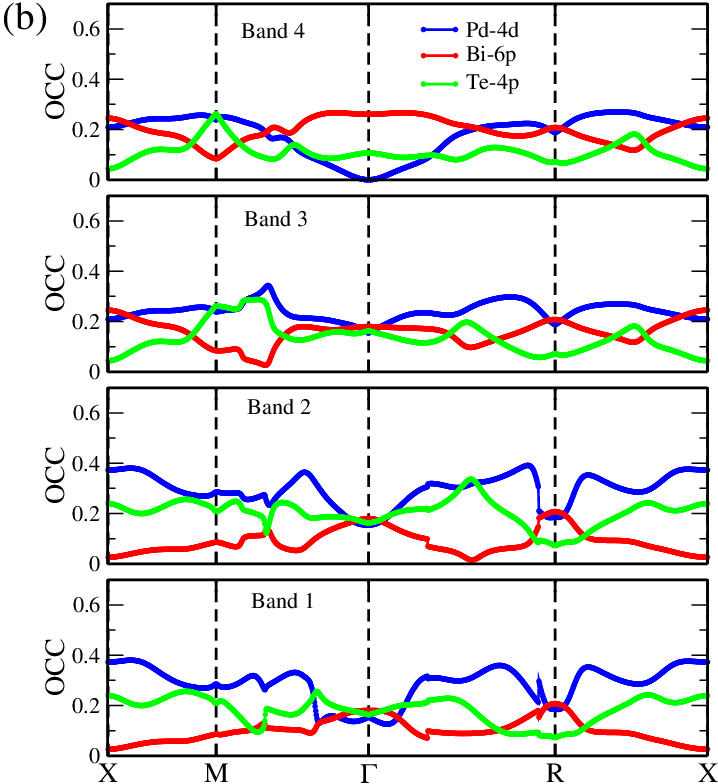}
\caption*{Fig. S4: \small Orbital character contribution to the bands without SOC (a) PdSbSe and (b) PdBiTe }
\end{figure*}

\section*{III.\hspace{0.3cm}With SOC}

\subsection*{A.\hspace{0.3cm}Electronic Structures}
\begin{figure*}[h] 
\includegraphics[width=0.48\linewidth]{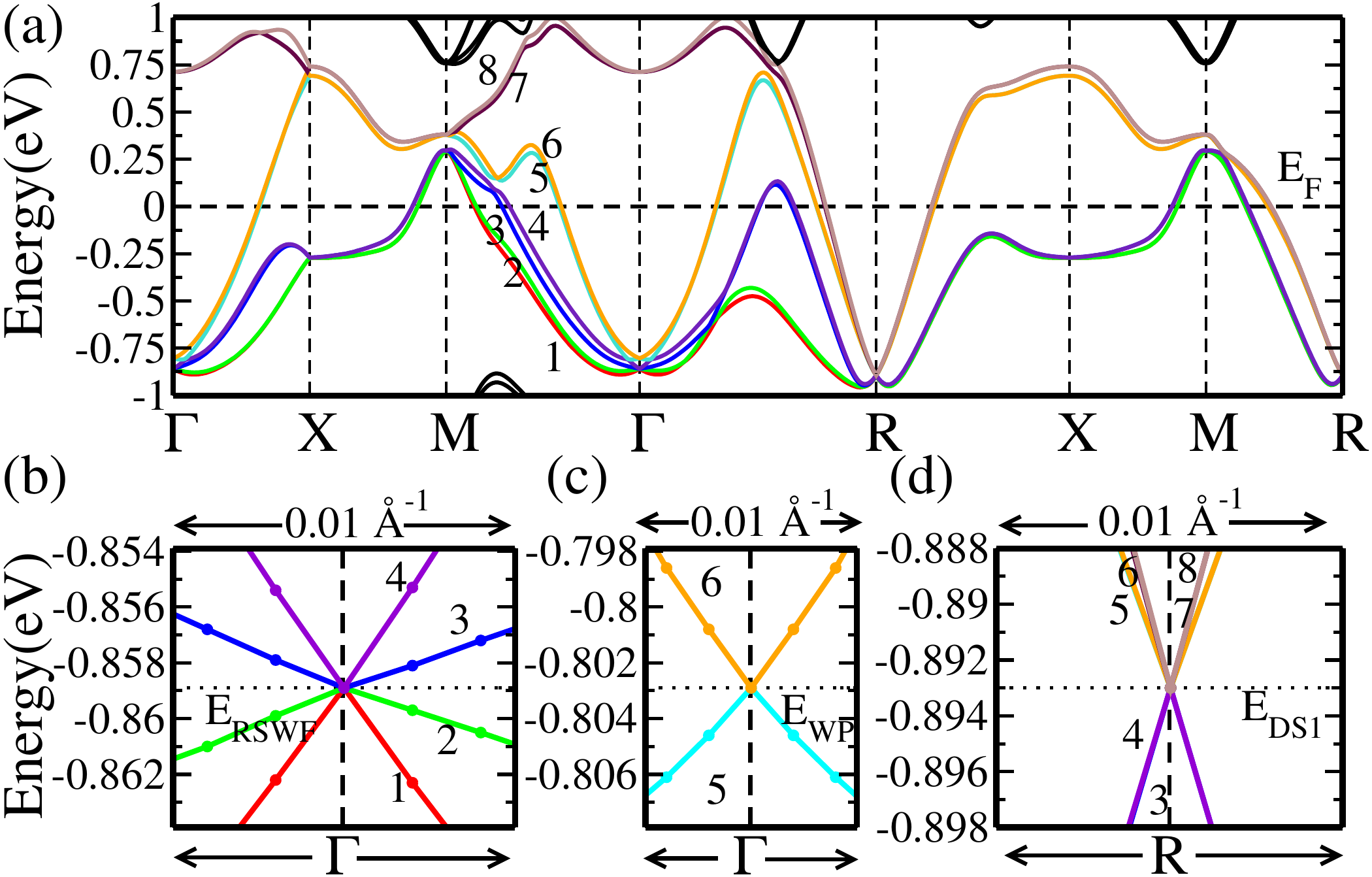}
\includegraphics[width=0.48\linewidth]{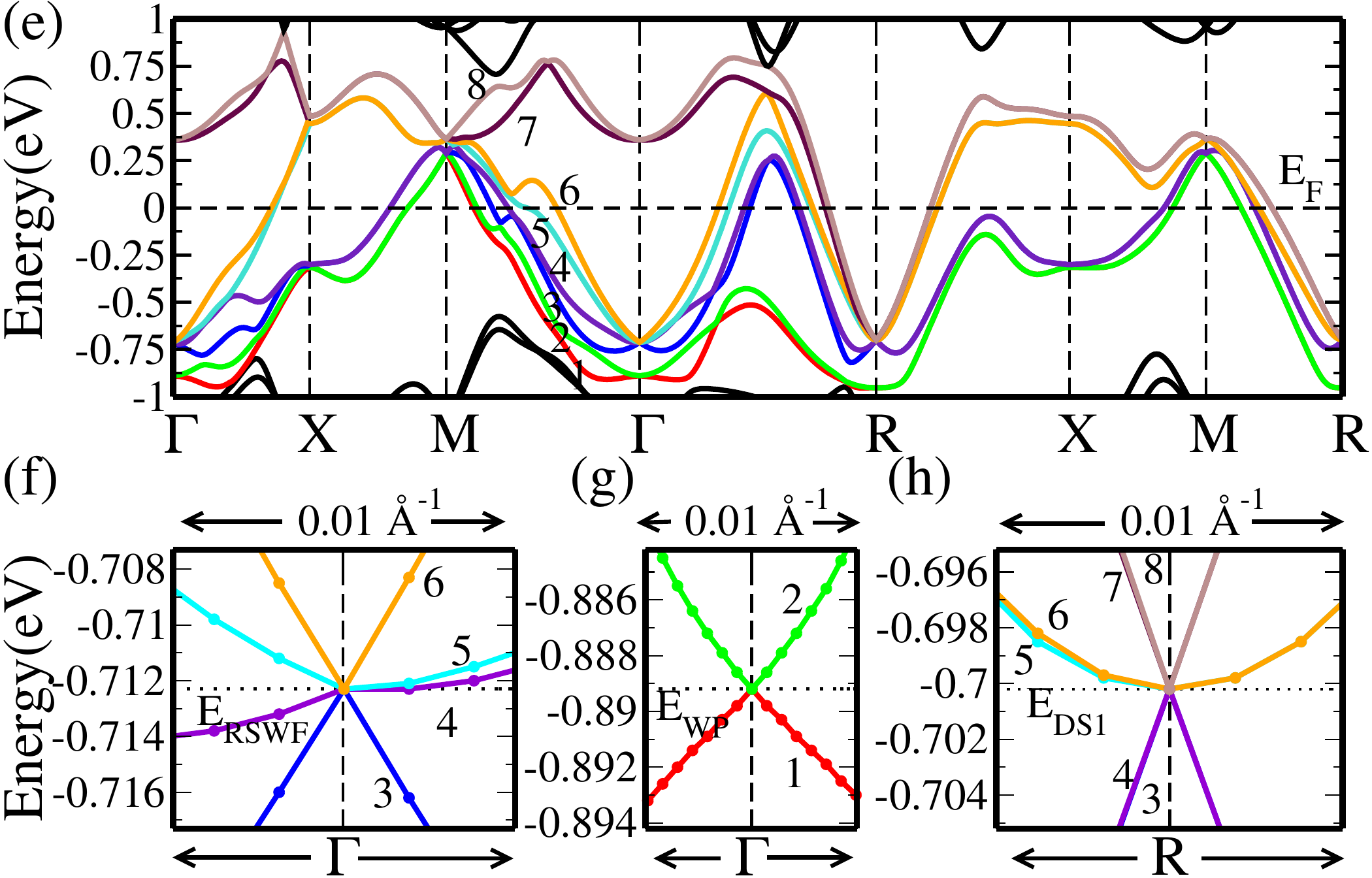}

\caption*{Fig. S5: \small Electronic structure, in presence of  SOC, of  (a) PdSbSe and (e)PdBiTe. Sub-figures (b) and (f) present enlarged views of the dispersion near $\Gamma$-point showing a Rarita-Schwinger-Weyl point (RSWP); panels (c) indicates a type-II WP along $\Gamma$-R and $\Gamma$-M;  (d) and (h) depicts the double spin-1 excitation at R-point. }
\end{figure*}
\newpage
\subsection*{B.\hspace{0.3cm}3-D Energy dispersion}
\begin{figure*}[h] 
\includegraphics[width=0.24\linewidth,height=3cm]{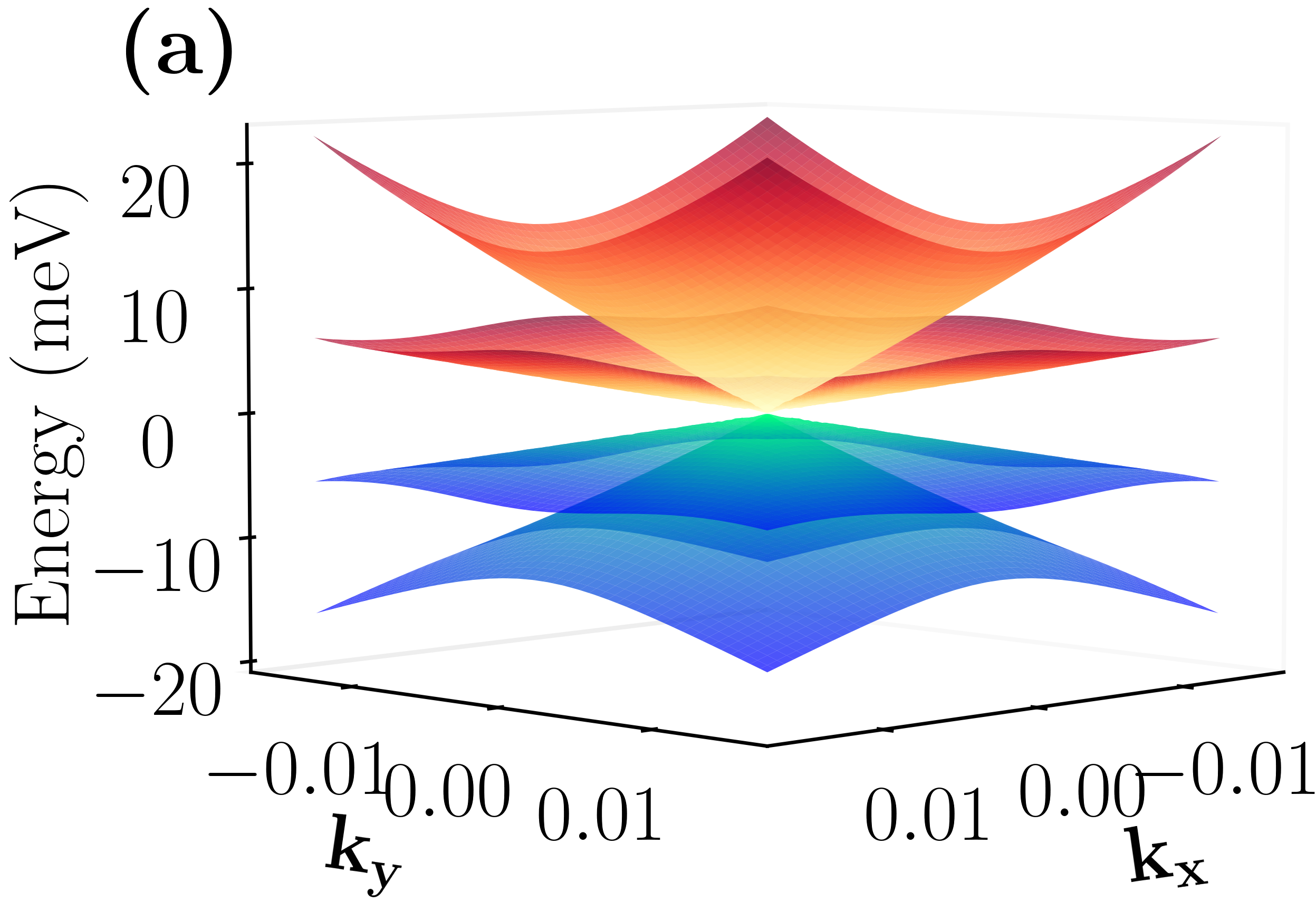}
\includegraphics[width=0.24\linewidth,height=3cm]{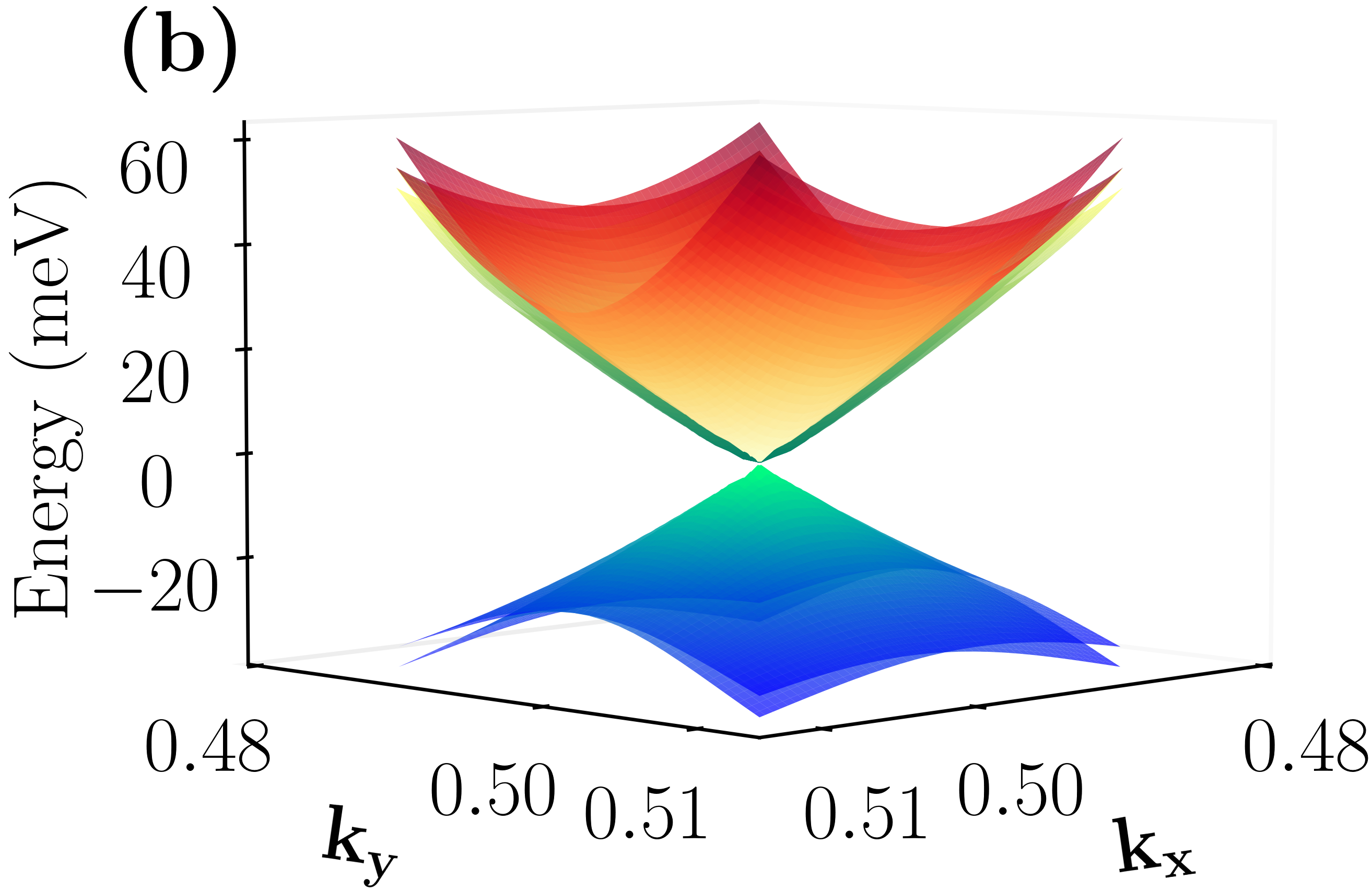}
\includegraphics[width=0.24\linewidth,height=3cm]{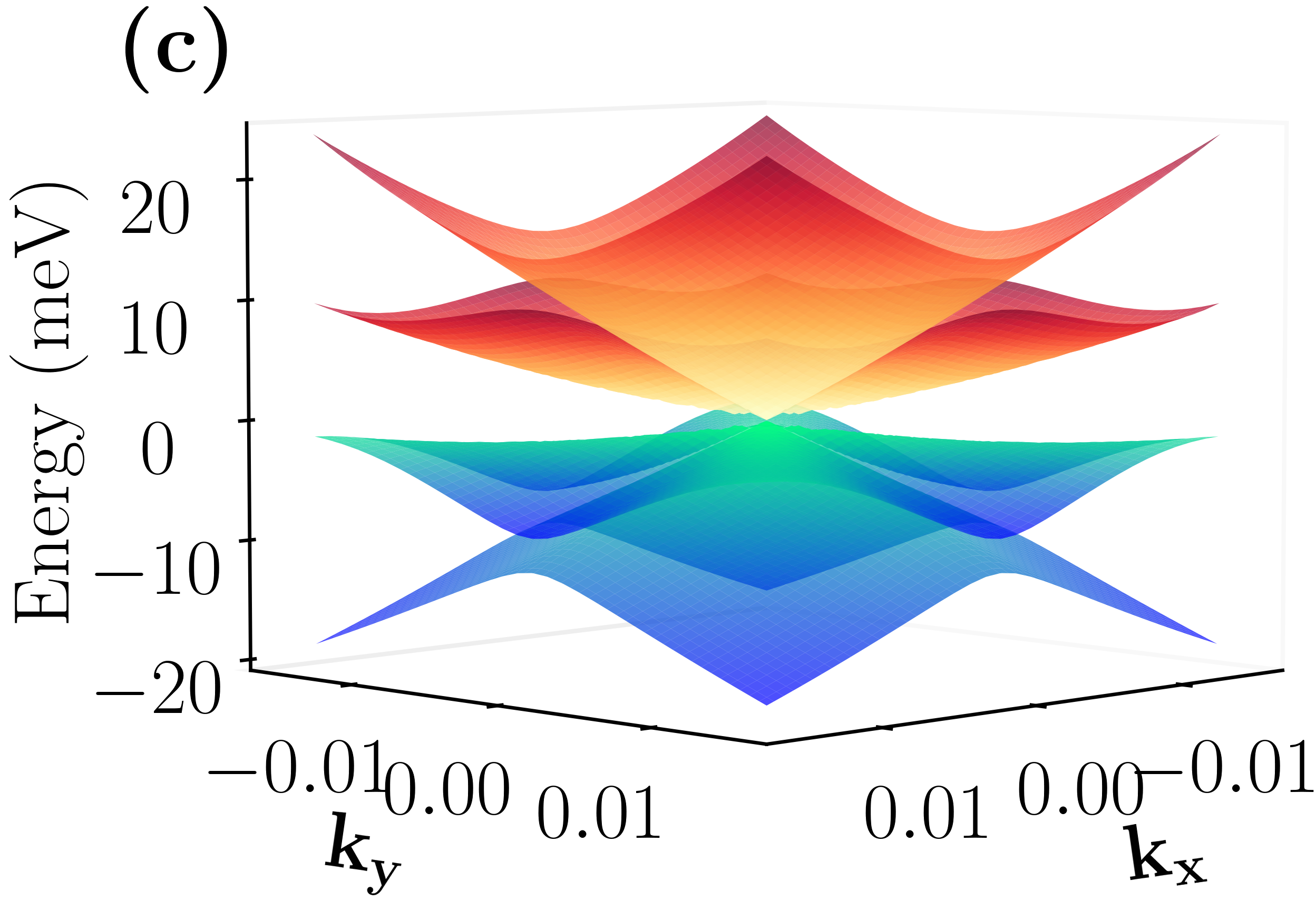}
\includegraphics[width=0.24\linewidth,height=3cm]{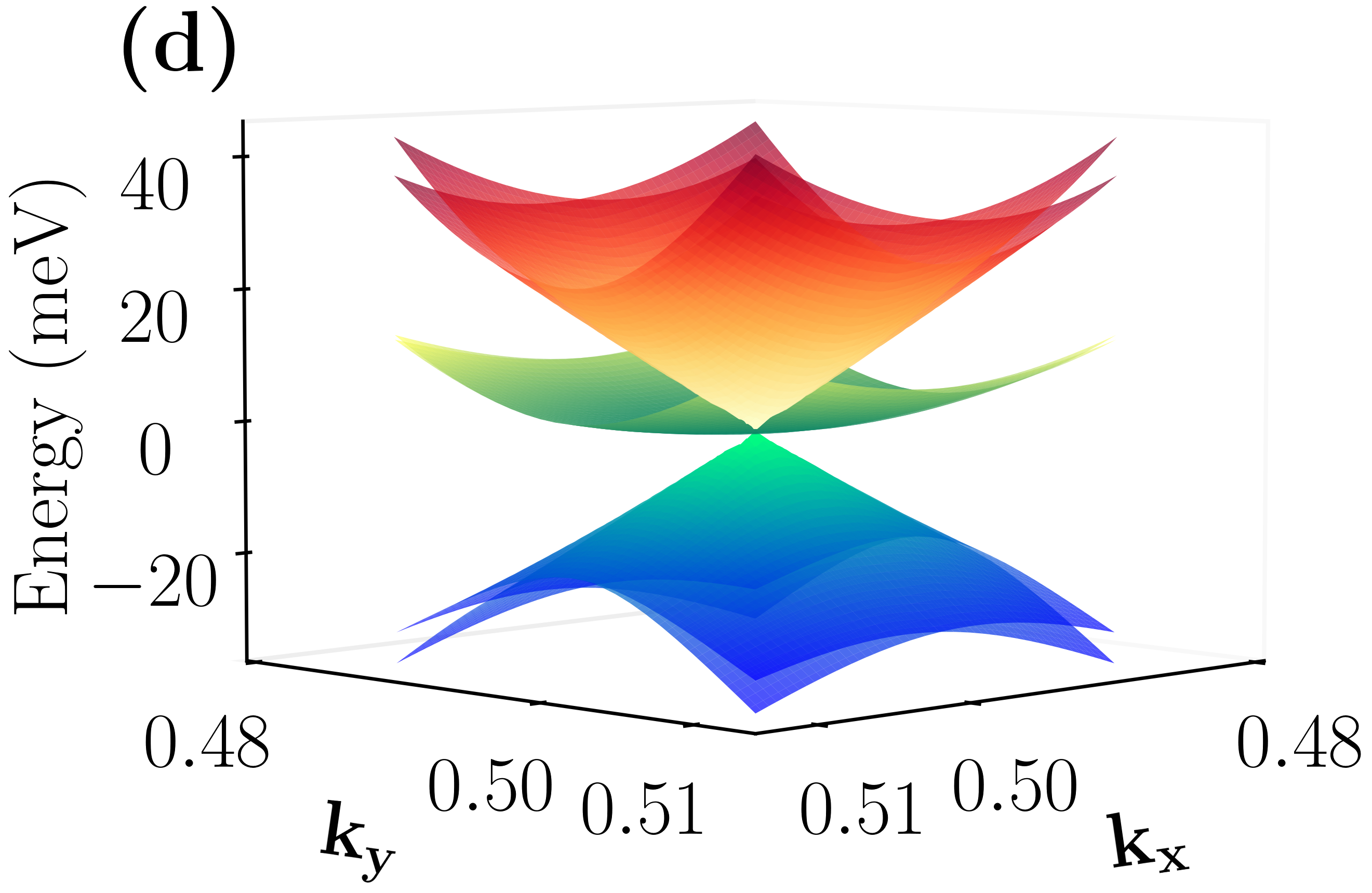}
\caption*{Fig. S6: Three-dimensional dispersion around high symmetric point $\Gamma$and R with SOC.  RSWP at $\Gamma$ in (a) PdSbSe and (c) PdBiTe. A double spin-1 excitation   at R-point in (b)PdSbSe and (d) PdBiTe.}
\end{figure*}

\begin{figure*}[h]
\includegraphics[width=0.32\linewidth]{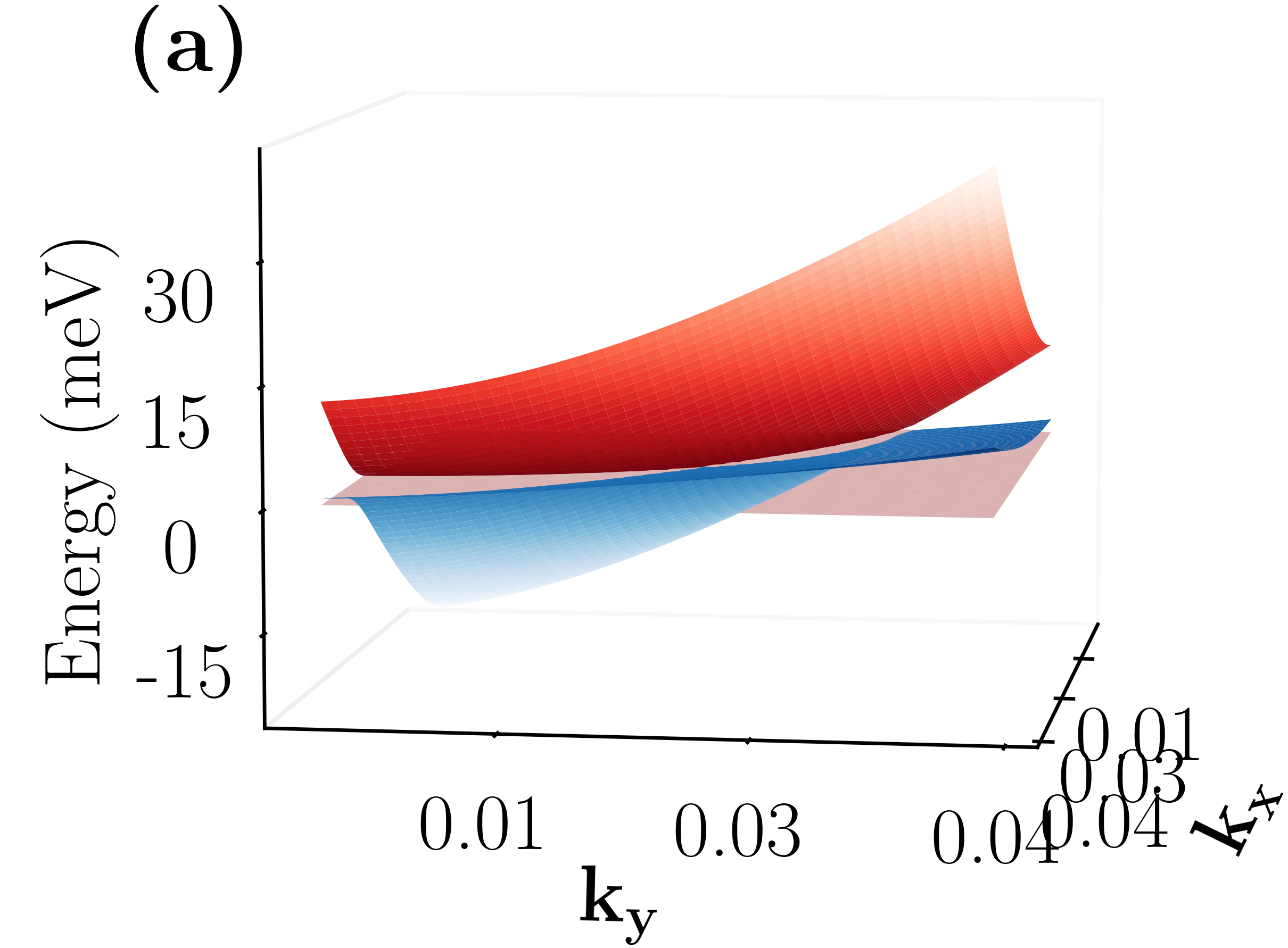}
\includegraphics[width=0.32\linewidth]{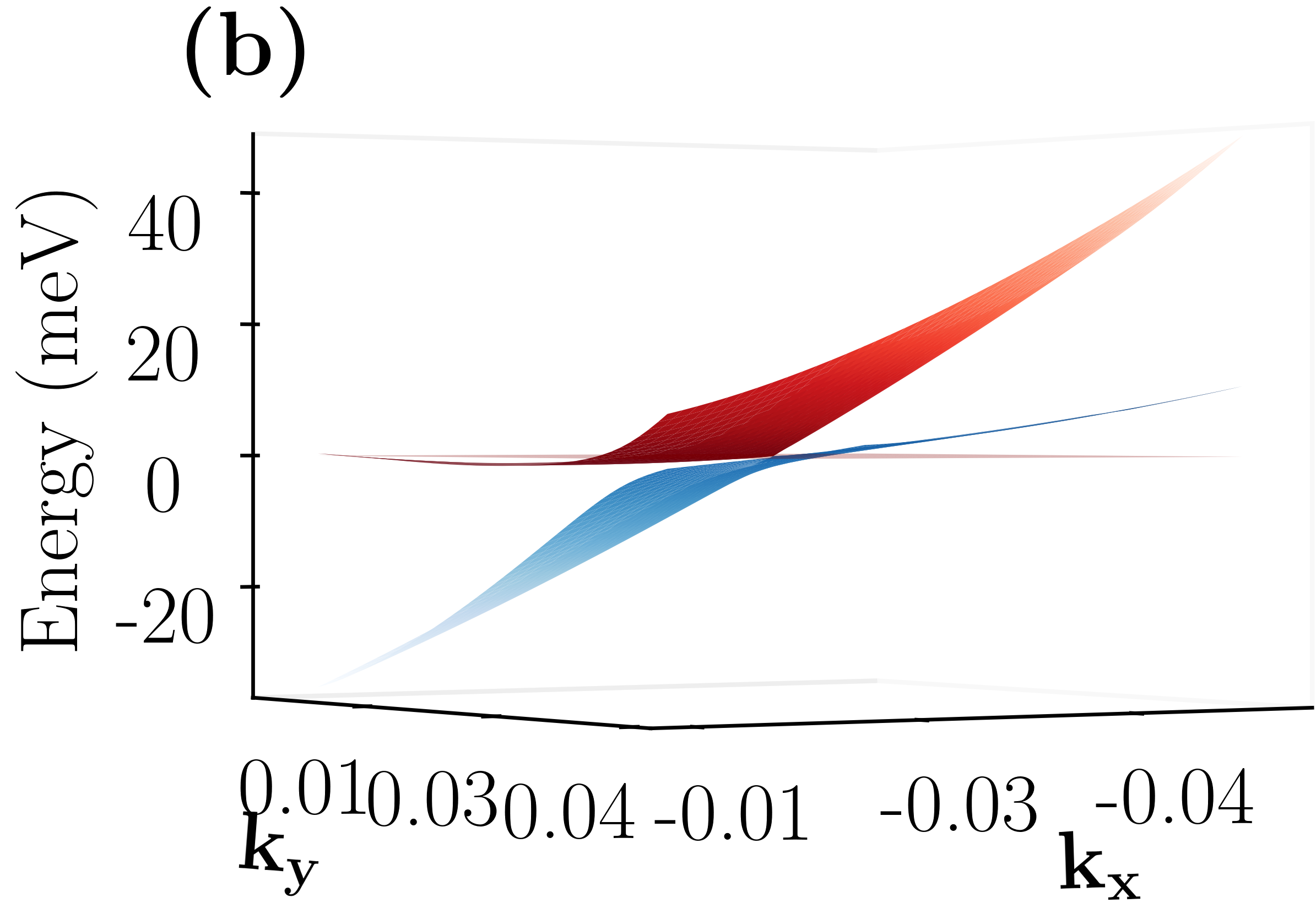}
\includegraphics[width=0.32\linewidth]{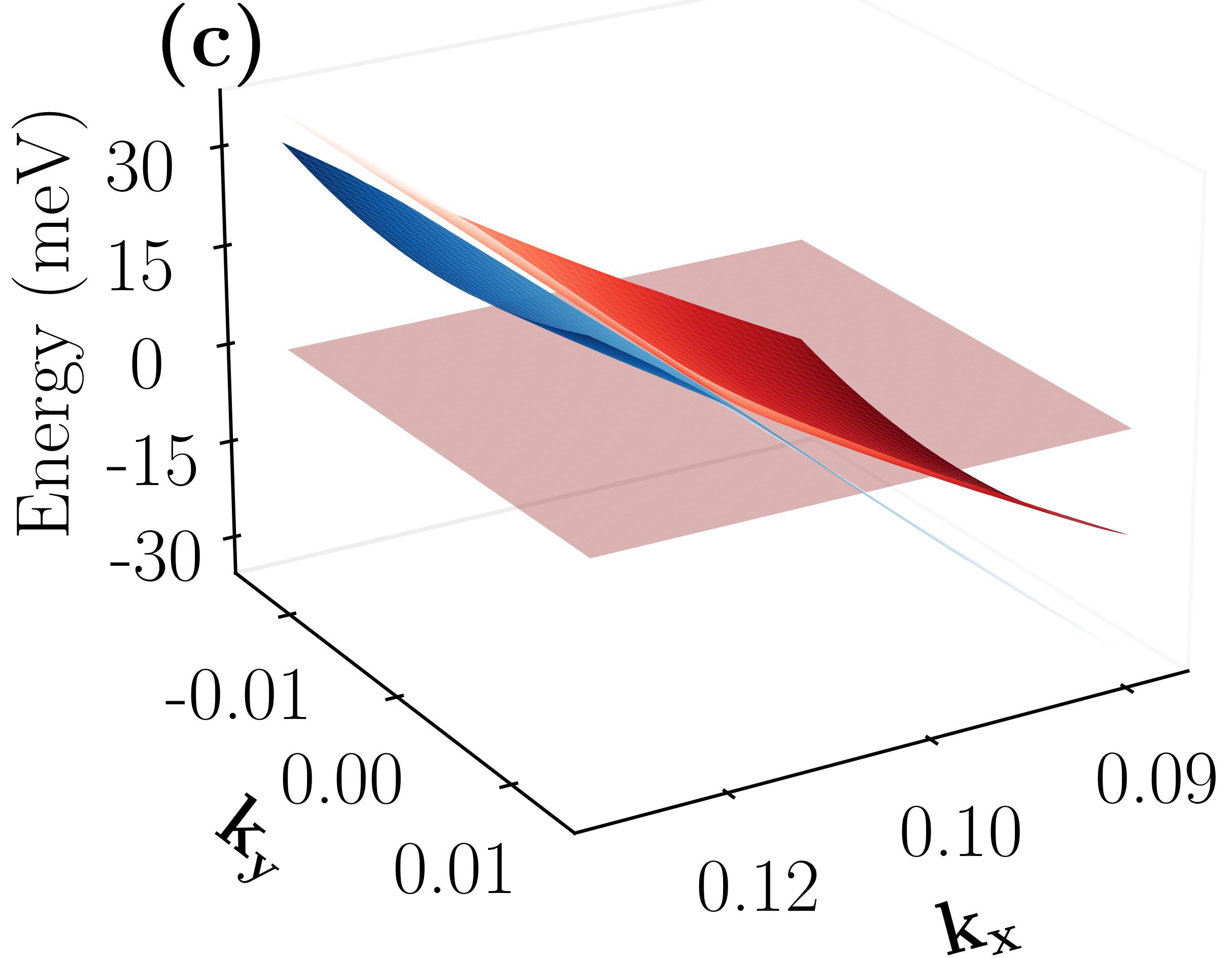}
\caption*{Fig. S7: Three dimensional energy dispersion around the obtained Weyl points in presence of SOC. The linearly dispersing bands from the Weyl point cross the isoenergy surface verifying the type-II behavior of the Weyl points of these crossings. The Weyl point energy is set to zero and band energies are scaled accordingly. The grey plane represents the zero energy. 3-D dispersions are as follows :  for PdSbSe (a) along $\Gamma$-R at point (0.024, 0.024, 0.024)$\frac{2\pi}{a}$, (b) at position (-0.035, 0.024, 0)$\frac{2\pi}{a}$ ; (c) PdBiTe at at position (0.108, 0, 0.019)$\frac{2\pi}{a}$ }
\end{figure*}
\subsection*{C.\hspace{0.3cm}Weyl point coordinates}

\begin{table}[h]
\small
\caption*{\label{tab_lattice_parameter}%
Table ST3:  The coordinates of Type-II WP in PdAsS, PdBiTe and PdSbSe in presence of SOC. We also verified topological charges of these Weyl points found them to either of $\pm1$. }
\begin{ruledtabular}

\begin{tabular}{lcccc}
\textrm{Material}&
\textrm{{along $\Gamma$-R } }&
\textrm{{Energy (eV)}}&
\textrm{General positions}&
\textrm{{Energy (eV)}}\\ 
&($k_x,k_y,k_z$)$\dfrac{2\pi}{a}$&  Fermi energy $E_F$=0&($k_x,k_y,k_z$)$\dfrac{2\pi}{a}$& Fermi energy $E_F$=0 \\
\colrule
&&&&\\

PdAsS& (0.033,0.033,0.033),(-0.033,-0.033,0.033)&-0.4692698&(0.000,$\pm$ 0.031,$\pm$ 0.051)& -0.4681531 \\
&  (-0.033,0.033,-0.033),(0.033,-0.033,-0.033) &&($\pm$0.031, $\pm$0.051, 0.000)&\\
&   (-0.033,-0.033,-0.033),(-0.033,0.033,0.033)&& ($\pm$0.051, 0.000, $\pm$0.031)&\\

&    (0.033,-0.033,0.033),(0.033,0.033,-0.033)& &&\\
PdSbSe	& (0.024, 0.024, 0.024),(-0.024, -0.024, 0.024) &   -0.8089258& ($\pm$0.024, 0.000, $\pm$0.035) &-0.8088427 \\ 		
&(-0.024, 0.024, -0.024),(0.024, -0.024,-0.024)&&(0.000, $\pm$0.035, $\pm$0.024) &\\

& (-0.024,-0.024,-0.024),(-0.024,0.024,0.024)& & ($\pm$0.035, $\pm$0.024, 0.000) &\\
&(0.024,-0.024,0.024),(0.024,0.024,-0.024)&&&\\

PdBiTe& -&-&($\pm$0.108, 0.000, $\pm$0.019) & -0.4681531 \\

&  -&-&(0.000, $\pm$0.019, $\pm$0.108)  &\\

& - &-&($\pm$0.019, $\pm$0.108, 0.000) &\\

\end{tabular}
\end{ruledtabular}
\end{table} 
\clearpage
\subsection*{D.\hspace{0.3cm}Orbital character plot}

\begin{figure*}[h] 
\includegraphics[width=0.48\linewidth,height=12.2cm]{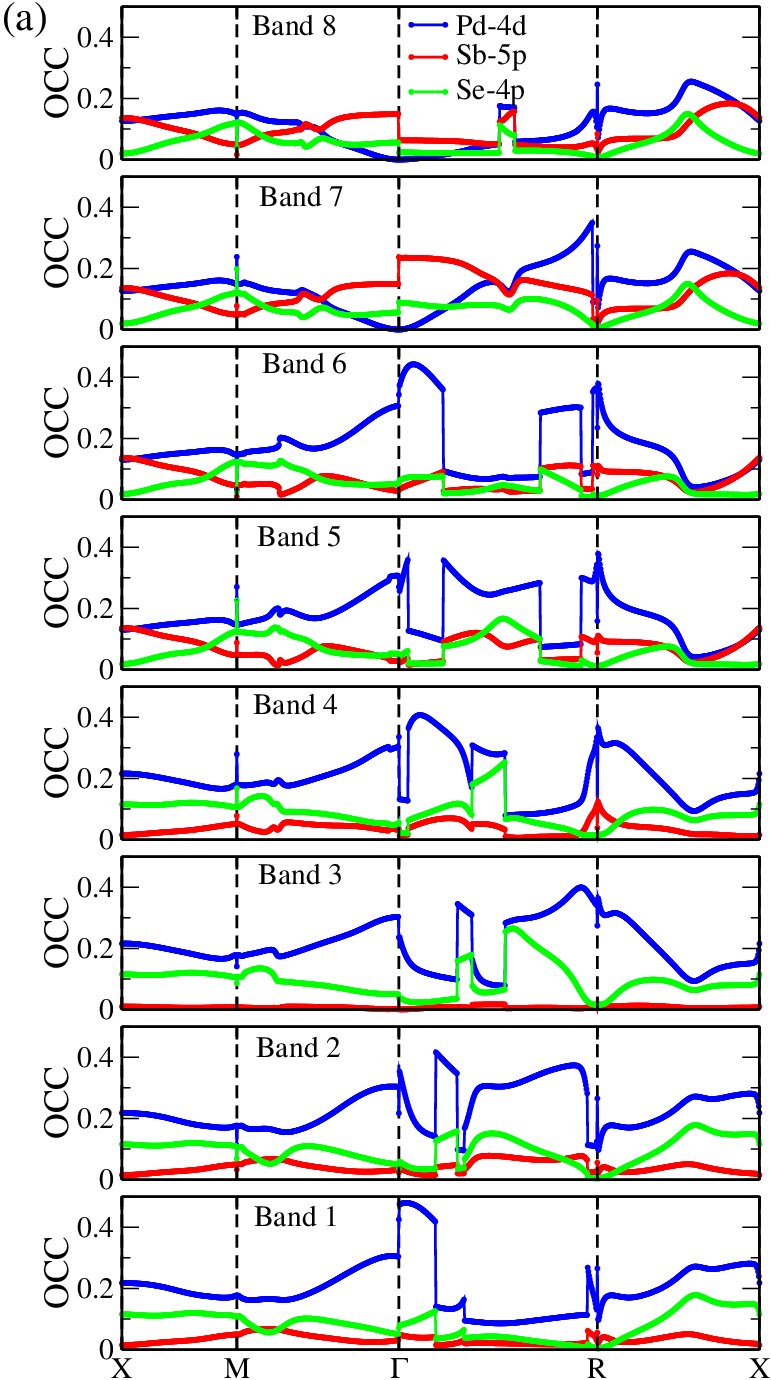}
\includegraphics[width=0.48\linewidth,height=12.2cm]{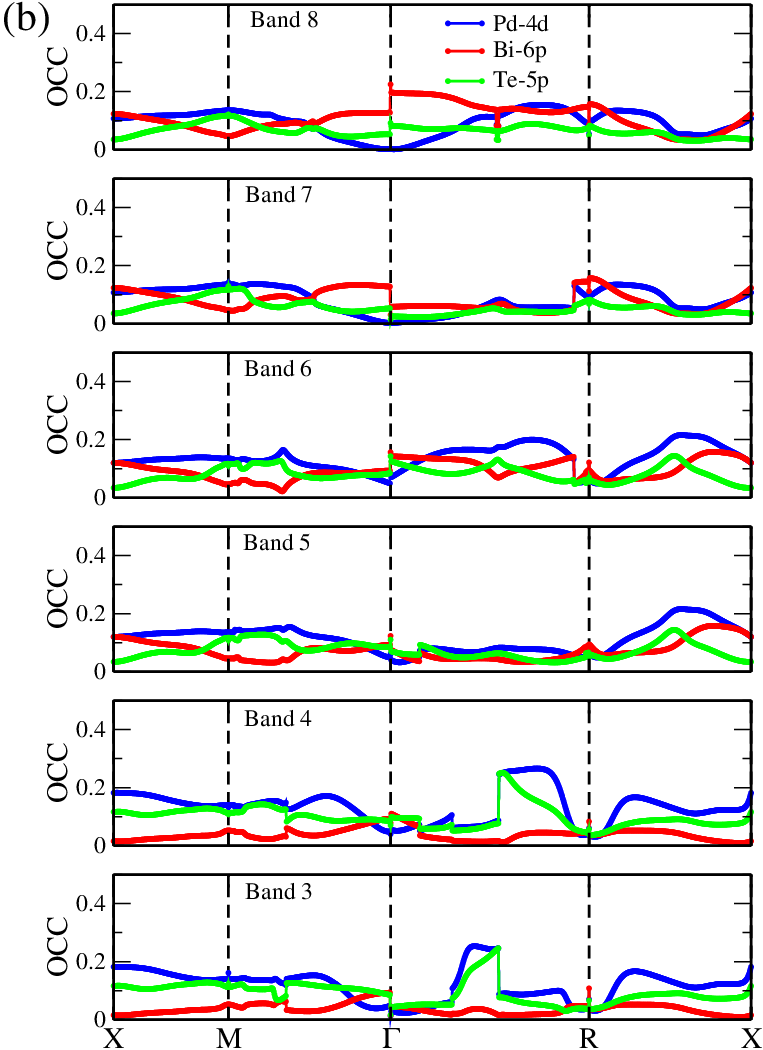}
\caption*{Fig. S8:\small Orbital character contribution to the bands with SOC (a) PdSbSe; here, RSWP is formed by Band 1-4 whereas the double spin-1 excitation is formed by Band 3-8 and (b) PdBiTe; RSWP (Band 1-4) and double spin-1 excitation (Band 1-6) }
\end{figure*}
\subsection*{E.\hspace{0.3cm}Fermi arcs}
\begin{figure*}[h] 
\includegraphics[width=0.98\linewidth,height=5cm]{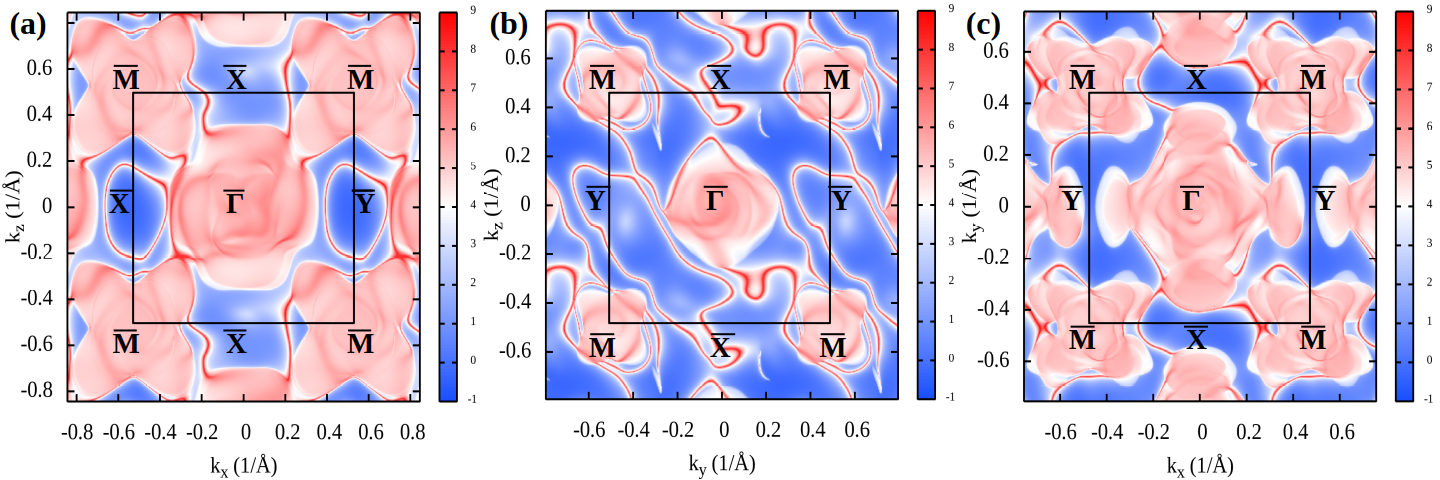}	
\caption*{Fig. S9:\small Constant energy plot of surface BZ of (a) PdAsS on (010) surface at energy -0.5 eV; (b) PdSbSe on surface (100) and energy 0.76 eV; (c) PdBiTe on surface (001) and energy -0.71 eV }
\end{figure*}

\end{document}